\documentclass[preprint]{aastex}
\usepackage{graphicx}

\newbox\grsign \setbox\grsign=\hbox{$>$} \newdimen\grdimen \grdimen=\ht\grsign
\newbox\simlessbox \newbox\simgreatbox \newbox\simpropbox
\setbox\simgreatbox=\hbox{\raise.5ex\hbox{$>$}\llap
     {\lower.5ex\hbox{$\sim$}}}\ht1=\grdimen\dp1=0pt
\setbox\simlessbox=\hbox{\raise.5ex\hbox{$<$}\llap
     {\lower.5ex\hbox{$\sim$}}}\ht2=\grdimen\dp2=0pt
\def\simgreat{\mathrel{\copy\simgreatbox}}
\def\simless{\mathrel{\copy\simlessbox}}

\begin{document}

\title{A Wavelet-Based Algorithm for the Spatial Analysis of Poisson Data}

\slugcomment{
27 August 2001.  To appear in {\it Ap. J. Supp.}}

\author{P.~E.~Freeman$^{1}$, V.~Kashyap$^{1}$, R.~Rosner$^{2}$, D.~Q.~Lamb$^{2}$}
\altaffiltext{1}{Harvard-Smithsonian Center for Astrophysics, 60 Garden Street, Cambridge, MA 02138}
\altaffiltext{2}{Department of Astronomy and Astrophysics, University
of Chicago, Chicago, IL 60637}

\begin{abstract}
Wavelets are scaleable, oscillatory functions that deviate from zero only
within a limited spatial regime and have average value zero, and
thus may be used to simultaneously characterize the shape, location, 
and strength of astronomical sources.  But in addition to their use
as source characterizers, wavelet functions are rapidly gaining currency 
within the source detection field.  Wavelet-based source detection
involves the correlation of
scaled wavelet functions with binned, two-dimensional image data.
If the chosen wavelet function exhibits the property of vanishing moments,
significantly non-zero correlation coefficients will be observed only
where there are high-order variations in the data; e.g.,
they will be observed in the vicinity of sources.
Source pixels are identified by comparing
each correlation coefficient with its probability sampling distribution,
which is a function of the (estimated or {\it a priori}-known) background
amplitude.

In this paper, we describe the mission-independent, wavelet-based
source detection algorithm {\tt WAVDETECT}, part of the freely available
{\it Chandra Interactive Analysis of Observations (CIAO)} software
package.  Our algorithm uses the Marr, or ``Mexican Hat" wavelet function,
but may be adapted for use with other wavelet functions.
Aspects of our algorithm include:
(1) the computation of local, exposure-corrected normalized
(i.e.~flat-fielded) background maps; (2)
the correction for exposure variations within the field-of-view
(due to, e.g., telescope support ribs or the edge of the field);
(3) its applicability within the low-counts regime, as it does
not require a minimum number of background counts per pixel for the
accurate computation of source detection thresholds;
(4) the generation of a source list in a manner that does not depend
upon a detailed knowledge of the point spread function (PSF) shape;
and (5) error analysis.
These features make our algorithm considerably more general than
previous methods developed for the analysis of
X-ray image data, especially in the low count regime.
We demonstrate the robustness of {\tt WAVDETECT} by applying it
to an image from an idealized detector with a spatially invariant Gaussian
PSF and an exposure map similar to that of the {\it Einstein} IPC; to
Pleiades Cluster data collected by the {\it ROSAT} PSPC;
and to simulated {\it Chandra} ACIS-I image of the Lockman Hole region.
\end{abstract}

\keywords{methods: data analysis --- techniques: image processing --- X-rays: general}

\section{Introduction}

\label{sect:intro}

The detection and characterization of astronomical sources 
becomes increasingly difficult as we attempt to observe 
these sources in the EUV, X-ray, and gamma-ray spectral regimes.
There are several reasons for this.
First, in these high-energy regimes,
source data may consist of only a few counts,
so that we must rely on the Poisson distribution when making
statistical inferences rather than using
Gaussian statistics that are considerably easier to apply,
but that are strictly applicable only
in the high-counts limit.
Second, spatially extended sources, such
as supernova remnants and galaxy clusters, exhibit bright diffuse
emission at high energies which may overlap with
point sources, rendering the latter more difficult both to detect and
characterize.  And third,
the present generation of broad-band high-energy telescopes,
unlike optical telescopes,
have spatially non-uniform point spread functions (PSFs) as
an unavoidable by-product of their design.
For instance, the PSF of the Wolter I-type High Resolution Mirror
Assembly (HRMA) on the {\it Chandra X-ray Observatory} (CXO)
has a 50\% encircled energy radius that varies
in width from $\approx$ 0.3$^{\arcsec}$ 
on-axis to $\simgreat$ 10$^{\arcsec}$ near the outer edges of an
ACIS-I chip ($\approx$ 10$^{\arcmin}$ off-axis).

A standard method for the analysis of Poisson count data involves the
application of the so-called ``sliding cell''
(Harnden et al.~1984).\footnote{
The method used, e.g., in the
{\it CIAO} source detection routine {\tt CELLDETECT}.} 
In sliding-cell analysis,
two co-aligned but differently sized square cells are placed at each
image pixel, with the number of counts in the 
annular region between cells providing an estimate
of the local background amplitude at the pixel.
This amplitude is then
used to compute the Poisson significance
of the observed number of counts in the inner cell.
The usefulness of a sliding-cell algorithm is limited both
when the field-of-view (FOV)
is crowded, since overlapping sources cannot be handled
and/or nearby sources contributing counts to the estimated background
may decrease detection sensitivity, 
and when sources are observed off-axis, since
uncertainties in the model of the PSF, which generally increase
with off-axis angle, can greatly affect source property
estimates (see, e.g., Kashyap et al.~1994).  
The sliding cell also may not provide accurate source property estimates
for extended sources.  Source characterization
can be improved by fitting the detected sources using maximum likelihood
methods (as in the algorithm of Hasinger, Johnston, \& Verbunt~1994), but 
the accuracy of this method is still limited by uncertainties in the PSF.

Within the last decade, astronomers have begun to apply wavelet functions
to the problem of source detection.
(For an introduction to the theory of wavelet functions, see, e.g.,
Mallat 1998 and references therein.)
These functions are scaleable, oscillatory, have a finite support
(i.e., are non-zero within a limited spatial regime),
and have an average value of zero;
they can be used to define a set of basis functions that act
as highly localized filters in both the spatial and frequency domains,
and thus are superior source characterizers.
Certain wavelet functions also exhibit the property of vanishing moments,
which is important for source detection: the integral of
the product of a wavelet with $N$ vanishing moments and a polynomial
of degree $\leq N-1$ is zero.  Thus the correlation of a suitably
chosen wavelet function with a photon counts image will yield 
correlation coefficients which are significantly large
only in the vicinity of localized high-order variations in the data,
e.g.~in the vicinity of astronomical sources, which appear as
PSF-broadened bumps with infinite power-series representation.
Wavelet-based source detection thus boils down to the statistical problem
of identifying
image pixels with ``significantly large" correlation coefficients.

Damiani et al.~(1997a) were the first to present a wavelet-based
generalized method for source detection and characterization, which they
apply to {\it ROSAT} PSPC data in a subsequent work (Damiani et al.~1997b).
Among its features, this method, unlike the others, uses exposure maps to 
handle situations in which, e.g., support rib shadows or the
edge of the FOV lie within the wavelet support,
allowing the analysis of the entire FOV.
Their algorithm is, however, instrument-dependent in that
they must make significant changes to it in order to account for
qualitative differences between exposure maps of different detectors
(e.g.~between the exposure maps of the {\it ROSAT} PSPC and {\it ROSAT}
HRI; Micela et al.~1999,
provide a short description of the changes that are necessary
to perform HRI image analysis).
In addition, while one can apply the Damiani et al.~method 
to data from other photon-counting detectors, the PSFs must be
very nearly Gaussian.
Last, they publish a function for the computation of
source detection thresholds which is strictly valid
only when the mean background amplitude per image pixel is
$\simgreat \frac{0.1~{\rm cts~pix}^{-1}}{\sigma^2}$, where 
$\sigma$ is the wavelet scale size.

In this paper (and in Freeman et al.~2001)
we describe our own algorithm for wavelet source detection and
characterization that has been developed
for a generic detector and implemented in the 
{\it Chandra Interactive Analysis of Observations (CIAO)}
routine
{\tt WAVDETECT}.\footnote{
The {\it CIAO} software package may downloaded from
{\tt http://asc.harvard.edu/ciao/}.
{\tt WAVDETECT}
is composed of {\tt WTRANSFORM}, a source detector, and {\tt WRECON},
a source list generator; these programs may be run separately.}
In subsequent papers (e.g.~Kashyap et al.~2001, in preparation) we will
discuss the application of this algorithm to
specific scientific problems.
Our algorithm is considerably more general and flexible than others that
have been developed, in that it can:
(1) operate effectively in the low
background counts regime, which is crucial because 
of the expected low particle and cosmic background count rates
for the {\it Chandra} detectors (the overall rate being
$\sim$ 10$^{-6}$ and 10$^{-7}$ ct sec$^{-1}$ pix$^{-1}$
for the {\it Chandra} ACIS and HRC detectors, respectively); 
and (2) operate effectively
regardless of the PSF {\it shape}, also crucial because of the
(non-Gaussian) nature of the off-axis {\it Chandra} PSFs.\footnote{
While it can operate to a limited extent if nothing at all
is known about the PSF,
our algorithm is most effective if characteristic PSF sizes, e.g.~the radii 
of circles containing 50\% of the encircled energy for different 
off-axis angles, are computable.}
It also (3) corrects for the effect of exposure variations in a general,
non-detector-specific manner.
Thus our algorithm may be immediately
adapted for the analysis of data from virtually any
other photon-counting detector.

In {\S}\ref{sect:wave}, we 
provide a brief description of wavelet functions, and define 
the Marr, or ``Mexican Hat" (MH) wavelet function which we use
in our algorithm.\footnote{
We note that while our algorithm
uses the MH function exclusively,
one can adapt our algorithm to work with other wavelet functions.}
We then present a
simple example in which we apply the MH function
to idealized data, in order to build the reader's intuition about 
how to interpret the results.
The MH function has been used often in astronomical wavelet 
analyses,\footnote{
However, we must note that 
the Mexican Hat is not the only wavelet function used by astronomers;
for instance, Rosati et al.~(1995) and Vikhlinin et al.~(1998)
use the Morlet wavelet function to
detect and analyze X-ray clusters in {\it ROSAT} PSPC images,
while Slezak, Durret, \& Gerbal (1994), Biviano et al.~(1996),
and Pierre \& Starck (1998) use the so-called ``B-splines" 
in their cluster analyses.
Also, there is the {\it \`a trous} algorithm of
Starck, Murtagh, \& Bijaoui 
(1995), Starck \& Murtagh (1998), and Starck \& Pierre (1998).
Methods based on this algorithm are similar
to the one which we describe in this paper; however,
because their use is limited to particular problems,
these methods are not sufficiently generalized to
be applied to the full-FOV data of an arbitrary detector.
For instance, these methods generally 
do not take into account exposure variations within the FOV, or
are applied only in limited regions not greatly affected by vignetting,
and they also generally ignore local variations in the background.}
in for example:
the analysis of $^{\rm 13}$CO spectral maps
of the molecular cloud L1551 (Gill \& Henrikson 1990; see
also Langer, Wilson, \& Anderson 1993, who apply Laplacian pyramid
transforms to a $^{\rm 13}$CO image map of Barnard 5);
the analysis of galaxy cluster structure in optically
derived catalogs (Slezak, Bijaoui, \& Mars 1990); the detection
and localization of features in optical CCD images of galaxies (Coupinot
et al.~1992); the analysis of substructures observed
in {\it ROSAT} PSPC
images of the Coma Cluster (Vikhlinin, Forman, \& Jones 1994);
the examination of
{\it Einstein} HRI and {\it ROSAT} PSPC images of Abell
1367 (Grebenev et al.~1995); the detection of serendipitous X-ray
clusters in archival {\it ROSAT} PSPC data (SHARC; cf.~Ulmer et
al.~1995, Freeman et al.~1996, Nichol et al.~1997); and the
analysis and modeling of X-ray emission in Abell clusters
(Lazzati \& Chincarini 1998; see also Lazzati et al.~1998)
and stellar clusters (Damiani et al.~1997b).

In {\S}\ref{sect:alg} we describe the basic steps of our algorithm,
in which sources are detected and characterized.
The basic steps of {\it source detection} include:
the correlation of the wavelet function with the data to create
the correlation image;
the estimation of the local background (if necessary)
in each pixel; the computation
of the source detection threshold in each pixel;
accounting for the effects of exposure variations within the FOV
on the correlation value;
and the identification and ``cleansing" of source counts from the 
image.  Source count cleansing is done iteratively, first by analyzing the
raw data, then the first version of the cleansed data, etc., until the
background is estimated.  This final background is used to compute source
detection thresholds, which are subsequently used to 
identify putative sources.  {\it Source characterization} involves
the combination of outputs
from a number of wavelet scales, and the creation
of source cells on the initial image within which
source properties (positions, count rates, etc.) are estimated.

In {\S}\ref{sect:ver} we demonstrate the efficacy of the algorithm
by applying it to a large variety of cases, such as: an idealized image 
containing simulated extended and point sources,
with spatially invariant Gaussian PSF and an exposure map
similar to that of the {\it Einstein} IPC;
a 32-ksec 
{\it ROSAT} PSPC observation of the Pleiades cluster (cf.~Micela et al.~1996);
and a simulated 30-ksec {\it Chandra} ACIS-I 
observation of the Lockman Hole region
(T.~Gaetz, private communication).
We then describe the differences between our method and previously published
methods, especially that of Damiani et al.,
in {\S}\ref{sect:comp}, and provide
our summary and conclusions in {\S}\ref{sect:sum}.
In future works, our algorithm will be applied to the analysis of
specific scientific problems, e.g.~observed spatial variations 
in the diffuse X-ray background of the Pleiades Cluster
(see Kashyap et al.~1996 and Kashyap et al.~2001, in preparation).

\section{Wavelets and Source Detection}

\label{sect:wave}

\subsection{Wavelet Properties}

Wavelets can be used to filter an image at a given length-scale.\footnote{
For a general introduction to wavelets, see, e.g., Mallat 1998 and 
references therein, and Daubechies 1992.}
This can be seen
by considering the difference between two smoothed functions:
one formed by convolving an (arbitrary) function $f$ with a
real-valued, non-negative, infinitely differentiable
smoothing function $\phi \in C^{\infty}$ 
whose size is characterized by a
scale $\sigma$ and satisfies the condition $\int \phi =$ 1
(e.g., the Gaussian function), and one formed by the convolution of $f$ with
the same smoothing function with scale size $\sigma + d\sigma$
(Holschneider 1995).
Because all structure at length-scales smaller than the smoothing
scales would be suppressed, the
difference of these two smoothed functions will provide information
about the details of $f$ that are
introduced at the scale $\sigma$ itself.
In the situation of interest to us, the analysis of two-dimensional images,
the relationship between $\phi$ and the wavelet function $W'$ is,
in the limits $d\sigma_x,d\sigma_y \rightarrow$ 0,
\begin{equation}
W_m'(\frac{x}{\sigma_x},\frac{y}{\sigma_y})~=~-\left(\sigma_x \frac{\partial}{{\partial}\sigma_x} + \sigma_y \frac{\partial}{{\partial}\sigma_y}\right) \phi(\frac{x}{\sigma_x},\frac{y}{\sigma_y}) .
\end{equation}
(The reason for our use of a prime symbol will become apparent below.)
This is an analyzing wavelet, or mother wavelet (hence the subscript $m$).
Other members of the
same wavelet family (so-called ``atoms'')
can be generated from this mother wavelet via
dilations and translations:
\begin{equation}
W'(a,b;\sigma_x,\sigma_y;x,y) \equiv \frac{1}{\sigma_x\sigma_y} W_m'\left(\frac{x-a}{\sigma_x},\frac{y-b}{\sigma_y}\right) .
\label{eqn:trans}
\end{equation}
$\sigma_x$ and $\sigma_y$ are 
the dilation parameters, and $a$ and $b$ are
the translation parameters.

\begin{figure}
\begin{center}
\includegraphics[angle=-90,scale=0.75]{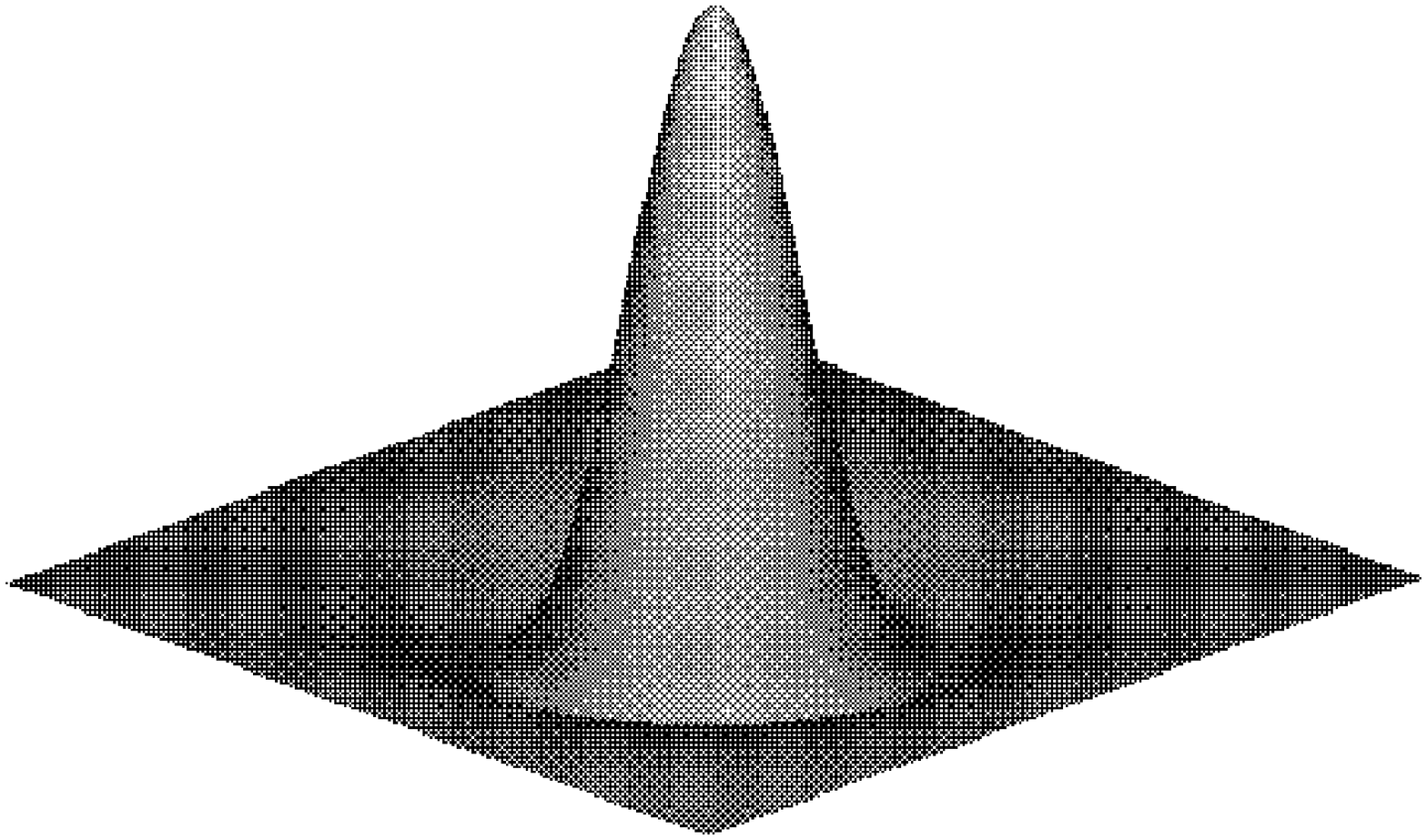}
\end{center}
\caption{
The two-dimensional Marr, or Mexican Hat, wavelet function
(eq.~\ref{eqn:wave}).
}
\label{fig:wave}
\end{figure}

There are many functions $\phi$ that can be used to create a mother
wavelet.  One example is the two-dimensional Gaussian function:
\begin{equation}
\phi(\frac{x}{\sigma_x},\frac{y}{\sigma_y})~=~\frac{1}{2{\pi}\sigma_x\sigma_y}\exp\left(-\frac{x^2}{2\sigma_x^2}-\frac{y^2}{2\sigma_y^2}\right) \,,
\end{equation}
from which is created the Marr, or ``Mexican Hat" (MH), wavelet function:
\begin{eqnarray}
W_m'(\frac{x}{\sigma_x},\frac{y}{\sigma_y})&=&\frac{1}{2{\pi}\sigma_x\sigma_y}\left[2-\frac{x^2}{\sigma_x^2}-\frac{y^2}{\sigma_y^2}\right]e^{-\frac{x^2}{2\sigma_x^2}-\frac{y^2}{2\sigma_y^2}} \label{eqn:wave1} \\
&=&\frac{1}{2{\pi}\sigma_x\sigma_y} W_m(\frac{x}{\sigma_x},\frac{y}{\sigma_y}) \,.
\label{eqn:wave}
\end{eqnarray}
This is the wavelet function we use in our source detection algorithm.
(Note that we use the function $W_m$ instead of $W_m'$,
to be consistent with Damiani et al.~1997a.)  The MH function
has a positive kernel with shape similar to a canonical PSF ($PW$),
surrounded by a negative annulus ($NW$; Figure \ref{fig:wave}).
Ellipses with semi-axes of length $\sqrt{2}\sigma_{x;y}$, 2$\sigma_{x;y}$,
and 5$\sigma_{x;y}$ describe the boundary between 
$PW$ and $NW$, the minimum of the MH function,
and the effective support of the MH function, respectively.

The MH function offers several advantages which motivate its use for
source detection: (1) it has
two vanishing moments (i.e.~the correlation of the MH function
with constant or linear functions is zero), so that it acts to
suppress the contribution of the (generally spatially constant)
background to the correlation coefficients;
(2) while it does not have compact support, and cannot be used
to construct a set of orthogonal basis functions, a
dyadic sequence of MH functions (i.e., with MH functions with scales
separated by factors of two) is
sufficient to sample the entire frequency domain, because of
the limited extent of an MH function's Fourier transform;
(3) its limited extent in both spatial and Fourier domains helps
to minimize effects of aliasing;
and (4) it is analytically manipulable, so that many
numerical operations may be performed 
using analytically derivable functions
(see Appendix \ref{sect:mh}), which can reduce computation time significantly.

\subsection{A Simple Example}

\label{sect:simp}

Before describing our algorithm in detail, we present a simplified 
example to help build the reader's intuition.
We assume that we are analyzing a subset of an
evenly exposed image with the MH wavelet function,
far from any edge of the FOV.
Within this image, the counts in each pixel, $D_{i,j}$, are sampled
from a function which has a constant, relatively large ($\gg$ 1) amplitude,
which we denote $B$.
These assumptions allow us to build 
intuition with a minimum of unnecessary detail (such as
correcting for exposure variations, etc.), and should not be construed
as being reflective of limitations on the applicability of our algorithm.

As will be described in {\S}\ref{sect:alg}, the first step in our
algorithm is to correlate the wavelet function $W$ and the binned
image data $D$.  We denote correlation using the symbol
$<...{\star}...>$; in this case, we would write 
$C =\,<W{\star}D>$, or for a particular pixel,
$C_{i,j} =\,<W{\star}D>_{i,j}$.\footnote{
This notation deviates somewhat from that of Mallat (1998), in which
$C_{i,j}$ would be written $WD[i,j]$; however, we feel our
notation makes complicated expressions in the remainder of this work
more easily interpretable.}
Because the average value of the wavelet is zero and the data are 
sampled from a constant amplitude function, the mean correlation value 
will tend asymptotically to zero,
with statistical sampling causing positively and negatively valued deviations
from zero in individual pixels.
Indeed, for our simple situation, the resulting 
probability sampling distribution (PSD) of correlation values,
denoted $p(C \vert B)$, will tend asymptotically to a zero-mean
Gaussian, with width $\sigma_{G,C} \propto \sqrt{\sigma_x\sigma_yB}$
(Freeman et al.~1996, Damiani et al.~1996, Damiani et al.~1997a).

We now assume that there is a tightly bunched clump of counts in our
otherwise source-free subset image, which
could be caused by a Poisson sampling fluctuation or by an astronomical source.
We assume the clump has Gaussian shape with amplitude $A_G$ and
width $\sigma_G$, which
is comparable to the size of the PSF of the instrument.
The correlation of the MH function with a Gaussian
yields a MH function that has its centroid at the Gaussian centroid,
and has centroid amplitude $C_{\rm max}(\sigma_x,\sigma_y)$
proportional to $A_G$:
\begin{equation}
C_{\rm max}~=~\frac{A_G \sigma_x \sigma_y}{\sqrt{(\sigma_G^2+\sigma_x^2)(\sigma_G^2+\sigma_y^2)}} \left[2 - \frac{\sigma_G^2}{\sigma_G^2+\sigma_x^2} - \frac{\sigma_G^2}{\sigma_G^2+\sigma_y^2}\right] .
\label{eqn:cmax}
\end{equation}
For our simple example, $C_{\rm max} >$ 0, i.e.~$C_{\rm max}$ is always
greater than the mean of the PSD.  To determine whether we should
associated the clump with an astronomical source, we would compute
the integral of the PSD from $C_{\rm max}$ to $\infty$; if 
this quantity is smaller than a predefined significance threshold 
(e.g.~$<$ 10$^{-6}$, or $C_{\rm max} > 4.75\sigma_{G,C}$),
then we associate the clump with a source.

Because $\sigma_{G,C} \propto \sigma_x\sigma_y$, source detectability
will vary as a function of wavelet scale size.
In the limits $(\sigma_x,\sigma_y) \rightarrow$ 0 
and $\rightarrow \infty$,
the ratio
$\frac{C_{\rm max}(\sigma_x,\sigma_y)}{\sqrt{\sigma_x\sigma_y}}$
goes to zero, i.e.~the source becomes undetectable.
If we apply symmetric wavelets in our analysis,
the maximum value of this ratio
occurs at $\sigma_x = \sigma_y = \sqrt{3}\sigma_G$.
Hence sources are most easily detected when one analyzes the image
with a wavelet function that has a size similar to that of the source.
Since such a function acts to ``filter,"
or selectively enhance, structures of similar scale, this behavior
is to be expected.

If the clump is associated
with a previously known source, but is not detected, we can use
eq.~(\ref{eqn:cmax}) to determine the upper limit on
source counts, by substituting
the source detection threshold for $C_{\rm max}$, and solving for
$A_G$ (see, e.g., {\S}\ref{sect:ver1}).
(Because the Gaussian is normalized, $A_G$ is identically
the number of source counts.)  We stress that
the use of this method to place upper
limits on source counts is limited to cases where the PSF shape is
that of a two-dimensional Gaussian function, and should not be
used in the place of simulations if the PSF has arbitrary shape.

\section{Algorithm}

\label{sect:alg}

\subsection{Source Detection}

\label{sect:wtransform}

\begin{figure}
\begin{center}
\includegraphics[scale=0.75]{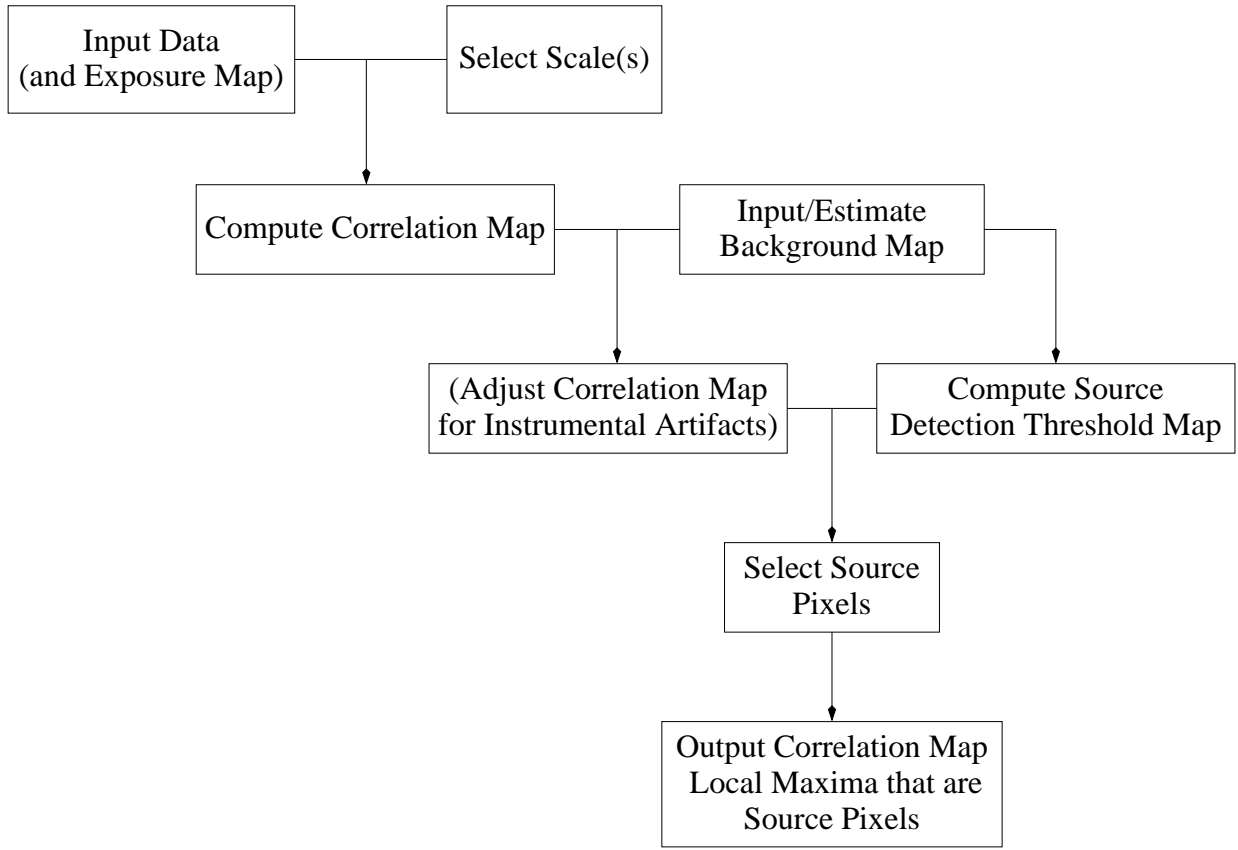}
\end{center}
\caption{
Flow chart illustrating the source detection algorithm
described in {\S}\ref{sect:wtransform}, as carried out
at selected scales $(\sigma_x,\sigma_y)$.
Parantheses indicate optional input or computations.  The
local background map may be estimated or provided by the
user; a flow chart illustrating how it can be iteratively estimated
using the input data (and exposure map) is given in 
Figure \ref{fig:bgflow}.
}
\label{fig:flow}
\end{figure}

Our first objective is {\it source detection}: the identification of
putative source pixels in binned, two-dimensional image data. 
This identification is normally done by carrying out the steps in 
the algorithm described below separately with each of a number of
wavelet functions (see also Figure \ref{fig:flow}).
The basic steps in this algorithm include (1)
the correlation of the data with the given wavelet function
({\S}\ref{sect:wcor}), and (2) the identification of image pixels
with correlation values larger than pre-defined thresholds for
source detection ({\S}\ref{sect:thresh}).
The second step requires knowledge of the local background in each
pixel (due to
unresolved point sources, diffuse astrophysical emission, particle
background, etc.).  If the background has not been determined
previously, then it must be estimated, e.g.~via the method described in
{\S}\ref{sect:back}.
Also, instrumental artifacts such as support rib structures, hot spots, and the
edge of the FOV can adversely affect the second step, 
resulting in the detection of instrumental features in the image,
which are astrophysically uninteresting.
In {\S}\ref{sect:wexp}, we describe a method for rejecting such
``instrumental sources."  Finally, in {\S}\ref{sect:werror} we describe
how the variances of the correlation and local background amplitudes
are estimated.

\subsubsection{Correlation of the Wavelet Function and the Data}

\label{sect:wcor}

The first step in the source detection process at a given scale is to
compute the correlation of the wavelet function 
$W(\sigma_x,\sigma_y;x,y)$
with the binned, two-dimensional image data $D$.\footnote{
In this section, we do not specifically refer to the Mexican Hat
wavelet to underscore the fact that our algorithm may be adapted for
use with other wavelet functions.}
The translation parameters $a$ and $b$ shown in eq.~(\ref{eqn:trans})
correspond to image pixel indices $i$ and $j$;
for a given pixel, the correlation value is
\begin{eqnarray}
C_{i,j}~&=&~\sum_{i'} \sum_{j'} W_{i-i',j-j'} D_{i',j'} \label{eqn:coranal} \\
&\equiv&~<W{\star}D>_{i,j} \,. \label{eqn:wd}
\end{eqnarray}
(Henceforth, if a quantity is given 
with subscripts, we are referring to its value at pixel $(i,j)$;
otherwise, we are referring to the quantity's array of values.)
The interested reader may find details about how $<W{\star}D>$ is computed 
in Appendix \ref{sect:mh}.

\subsubsection{Computation of the Source Detection Thresholds}

\label{sect:thresh}

To determine whether pixel $(i,j)$ should be associated with a
source, we compute the probability of observing the
correlation value $C_{i,j}$ if there are only background counts present
within the support of $W$:
\begin{equation}
S_{i,j}~=~\int_{C_{i,j}}^{\infty} dC p(C \vert B_{i,j}) \,.
\end{equation}
$S_{i,j}$ is dubbed the significance (or the Type I error; see, e.g.,
Eadie et al.~1971, pp.~215-216), and it is
the estimated probability that we would erroneously identify pixel
$(i,j)$ with a source.
In {\S}\ref{sect:simp}, we indicated how $S_{i,j}$ could be determined
analytically in the high-counts limit, where $p(C \vert B_{i,j})$ tends
asymptotically to a zero-mean Gaussian 
of width $\sqrt{2{\pi}\sigma_x\sigma_yB_{i,j}}$.  In general, however,
$p(C \vert B_{i,j})$ must be determined via simulations (see Appendix B).
Because the number of simulations we carried out is only sufficient
to directly determine significances $S_{i,j} \simgreat$ 10$^{-7}$, 
and because strong sources may have much greater significances
(in a qualitative sense), we instead compute a source detection threshold 
$C_{o,i,j}(S_o,B_{i,j})$ via the equation
\begin{equation}
S_o~=~\int_{C_{o,i,j}}^{\infty} dC p(C \vert B_{i,j}) \,,
\label{eqn:sigcalc}
\end{equation}
where $S_o$ is the user-specified threshold significance.\footnote{
One choice is $S_o = P^{-1}$, where $P$ is the number of analyzed pixels
in the image;
with this choice, the average number of false detections is one per image.}
If $C_{i,j} > C_{o,i,j}$, we associate the pixel $(i,j)$ with a source.

\subsubsection{Background Estimation}

\label{sect:back}

\begin{figure}
\begin{center}
\includegraphics[scale=0.75]{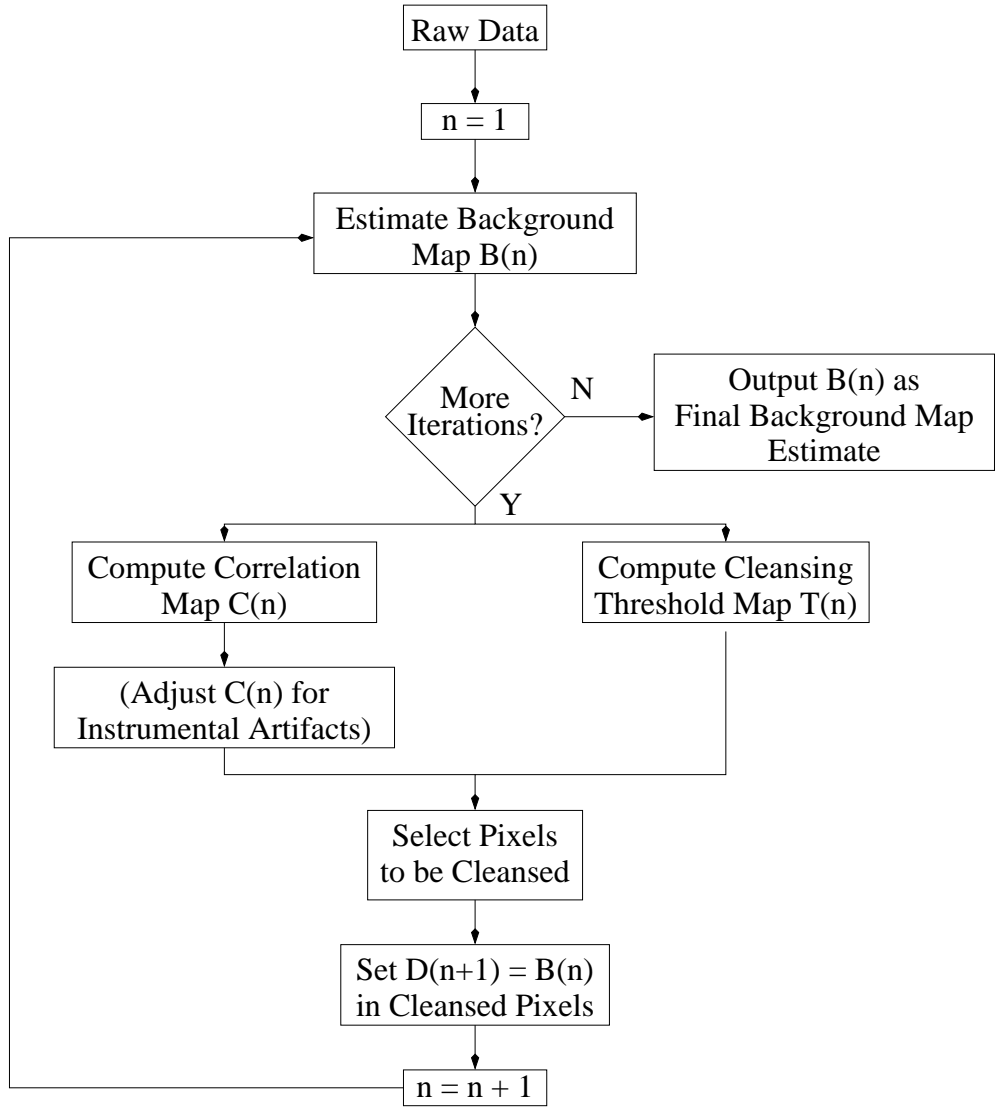}
\end{center}
\caption{
Flow chart illustrating the iterative local background estimation
method described in {\S}\ref{sect:back}.  Note that a background
map is output for each selected scale pair $(\sigma_x,\sigma_y)$.
Parantheses indicate optional computations.
}
\label{fig:bgflow}
\end{figure}

In this section, we describe how the local background counts
amplitude $B_{i,j}$ is estimated if it is unknown {\it a priori}
(see also Figure \ref{fig:bgflow}).
While there is no unique way to make this estimate, we seek a method
that does not depend on a detector's PSF, both for increased generality
and computational speed, and also because we want our algorithm to
be able to detect and analyze sources of arbitrary size, 
not just point sources.  We can fulfill this condition by
creating background maps {\it at each wavelet scale},\footnote{
These maps are later combined into a single map used in the calculation
of source properties.  See {\S}\ref{sect:nbkg}.}
using a localized function that is wavelet-scale, and not PSF-scale, dependent:
the wavelet negative annulus $NW(\sigma_x,\sigma_y)$ (i.e.,
$W$ with positive values reset to zero).  While the function $NW$ is
related to a wavelet function, {\it we stress that it itself is not a 
wavelet function, and that its use in background estimation
does not constitute a transform.}

If the exposure within the support of $NW$ is constant, and if 
$\sigma_x$ and $\sigma_y$ are sufficiently large such
that the integral of the source
counts distribution (i.e.~the PSF for a point source) over the $NW$
is insignificant with respect to the integrated background, then
one can estimate the background using the formula
\begin{equation}
B_{i,j}~=~\frac{\exp(1)}{4{\pi}\sigma_x\sigma_y} \vert <NW{\star}D>_{i,j}  \vert \,,
\label{eqn:bnoexp}
\end{equation}
where $4{\pi}\sigma_x\sigma_y{\slash}\exp(1)$ is the integrated volume
of $NW(\sigma_x,\sigma_y)$ (see Appendix \ref{sect:negann} for derivation).
If on the other hand the exposure is not constant, the exposure map\footnote{
If one does not provide an exposure map,
a flat one is assumed, to account for the edge of the FOV.}
can be used as a weighting function:
\begin{eqnarray}
B_{i,j}~&=&~E_{i,j} B_{{\rm norm},i,j} \nonumber \\
&=&~E_{i,j} \frac{<NW{\star}D>_{i,j}}{<NW{\star}E>_{i,j}} \label{eqn:back2} \\
&=&~E_{i,j} \frac{\sum_{i'} \sum_{j'} NW_{i-i',j-j'}D_{i',j'}}{\sum_{i'} \sum_{j'} NW_{i-i',j-j'}E_{i',j'}} \label{eqn:back} \,,
\end{eqnarray}
where $B_{{\rm norm},i,j}$ is the normalized (i.e.~flat-fielded) number of
expected background counts at pixel $(i,j)$.
We ignore the distinction between vignetted and non-vignetted components of the
background (e.g., the particle background) in our estimate, because of
degeneracy.  We note that if
the amplitude of the non-vignetted background
component $B^{\rm NV}$ is known, then one could in principle estimate
the background using a variation of eq.~(\ref{eqn:back}):
\begin{eqnarray}
B_{i,j}~&=&~E_{i,j} \frac{<NW{\star}(D-B^{\rm NV})>_{i,j}}{<NW{\star}E>_{i,j}} + B^{\rm NV} \nonumber \,.
\end{eqnarray}

There are three situations in which
counts from sources may bias the local background estimate:
(1) if the estimate is being made within diffuse extended emission;
(2) if the pixel in which the estimate is being made is a source pixel, but 
$\sigma_x$ and/or $\sigma_y$ is smaller than the source size $s$;
or
(3) if sources are located within the $NW$.

The first situation is a non-issue, because if the analysis goal is detection,
say of a source within a supernova remnant, then the diffuse
emission should be treated as a local background component: for
instance, should a clump of
counts be associated with a source, or with Poisson fluctuations
in the background {\it and} the diffuse emission?

\begin{figure}
\begin{center}
\includegraphics[scale=0.75]{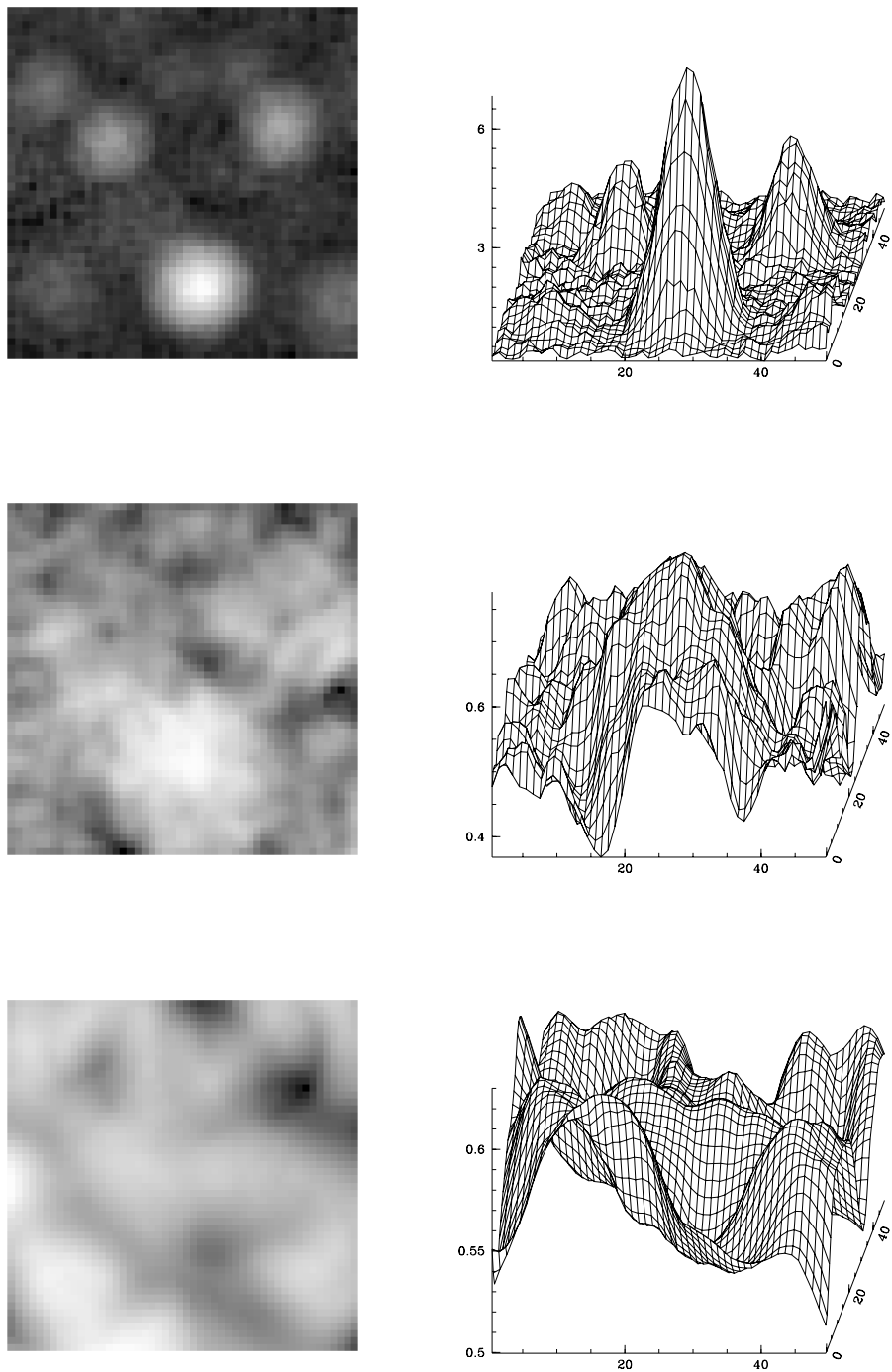}
\end{center}
\caption{
Illustration of how source counts may bias a final background
estimate ({\S}\ref{sect:back}), causing ``bumps" if the 
wavelet scale sizes $(\sigma_x,\sigma_y) \simless r_{\rm PSF}$,
the characteristic PSF size at a given pixel.
We use a 50$\times$50 subfield of the {\it ROSAT} PSPC Pleiades Cluster
image in which $r_{\rm PSF} \approx$ 3 pixels.
Top: image and surface plot of 
the background estimate for $\sigma_x$ = $\sigma_y$ = 1 pixel.
Middle: same as top, but for $\sigma_x$ = $\sigma_y$ = 2$\sqrt{2}$ pixels.
Bottom: same as top, but for $\sigma_x$ = $\sigma_y$ = 8 pixels.
}
\label{fig:bgbump}
\end{figure}

In the second situation, the background map will exhibit a ``bump" 
at the location of the source, whose amplitude is greatest where the
number of source counts within the support of the $NW$ is maximized,
generally at the source centroid (Figure \ref{fig:bgbump}).
Within the bump, the source detection threshold is overestimated, 
and thus this perhaps-otherwise-detectable source may remain
undetected {\it at the given scale}.  However, this is not a critical
problem, since if the source is detectable, it will be detected when
($\sigma_x$,$\sigma_y$) $\simgreat s$, a scale regime where
the bump is minimized, and the issue becomes
moot.  Note that these bumps do not adversely affect
source property estimation, since they are eliminated when the background
map that is used for source property estimates is computed 
({\S}\ref{sect:nbkg}).

\begin{figure}
\begin{center}
\includegraphics[scale=0.75]{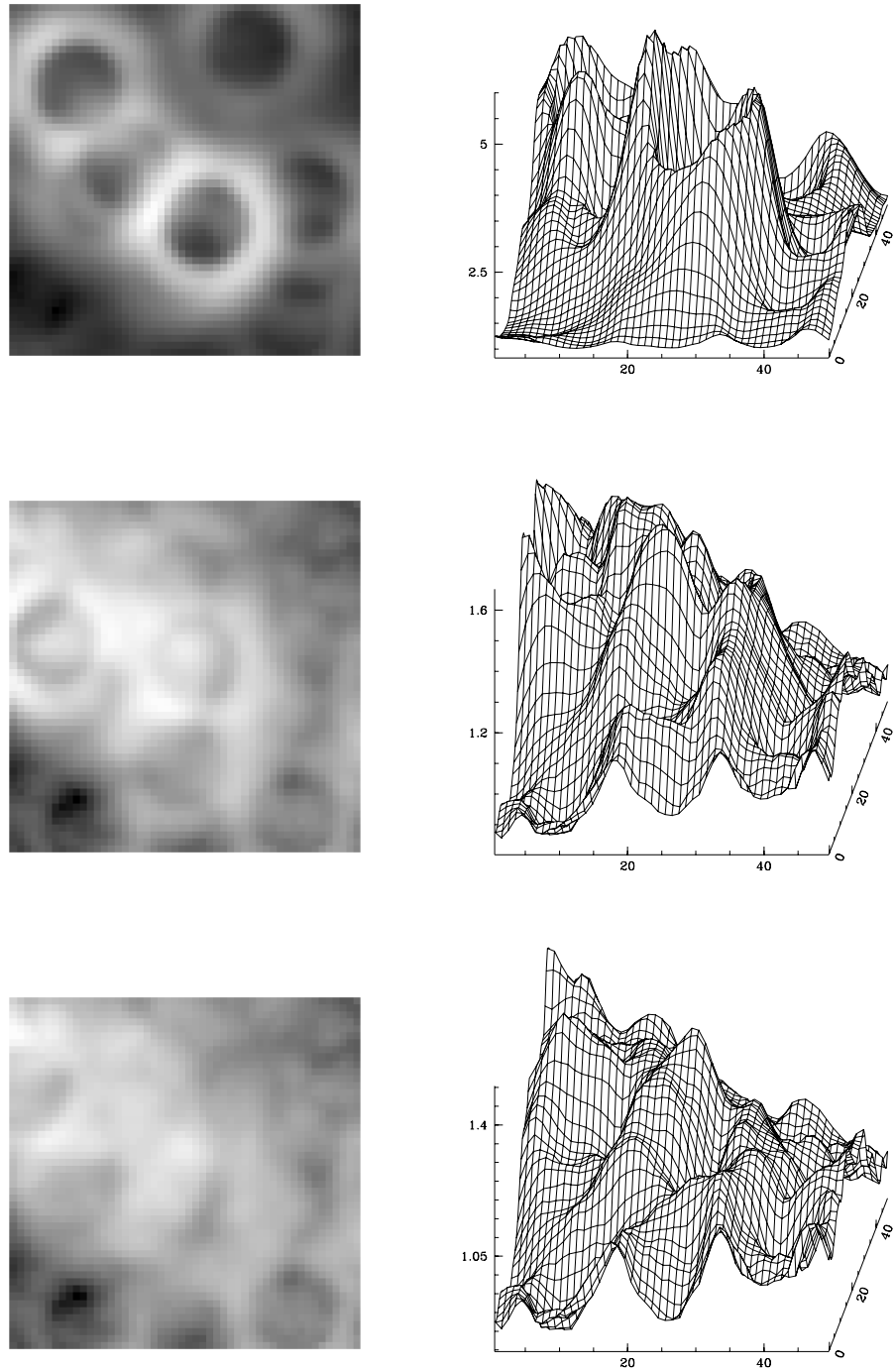}
\end{center}
\caption{
Illustration of how source counts may bias an initial background
estimate ({\S}\ref{sect:back}), causing ``rings" if sources are
located within the negative annulus of the wavelet.
We use a 50$\times$50 subfield of the {\it ROSAT} PSPC Pleiades Cluster image.
Top: image and surface plot of 
the background estimate after one iteration (i.e.~no ``cleansing").
Middle: same as top, but after two iterations.
Bottom: same at top, but after three iterations.
}
\label{fig:bgring}
\end{figure}

The third situation can be more problematic for source detection.
If there are sources in the FOV, then there will {\it always} be some
image pixels for which the background amplitude is overestimated:
for instance, for a symmetric wavelet function,
these pixels will surround sources in circular
``rings" of radius $\approx$ 2$\sigma$ (the radius at which $NW$ achieves
its minimum value; see Figure \ref{fig:bgring}).
In these rings, $C_o$ is overestimated, so that otherwise-detectable
sources whose locations coincide with these rings may go undetected.
Rings can appear regardless of the scale or source size,
and thus can impede source detection at all scales.
They can also adversely affect the computation of source properties, 
as a background that is overestimated in the vicinity of a source
can lead to underestimated count rates, etc., in the final source list.
(This last point is demonstrated below in Figure \ref{fig:spropest}, which
shows the effect of rings on the estimation of Pleiades Cluster source 
properties.  See {\S}\ref{sect:ver3}.)

One way to remove the rings is to remove source counts iteratively 
from the raw data, via the following algorithm: 
\begin{enumerate}
\item identify pixels to be ``cleansed" using $p(C \vert B)$ and the initial
background map, which we will dub $B_1$; 
\item mask out these pixels
or replace their data with other values, creating a new image
we denote $D_2$; 
\item estimate $B_n(D_n)$, where $n$ is the iteration
number ($n \geq$ 2); and 
\item if the background map is to be refined
yet again, determine $C_n(D_n)$, identify pixels to be cleansed
using $p(C \vert B_n)$, and return to step (2).  
\end{enumerate}
Then, when the final
background map $B_{\rm final}$ is determined, compare the original
values $C$ with $p(C \vert B_{\rm final})$ to create a final list
of putative source pixels.

Regarding steps (1) and (4), since the goal of this iterative approach
is to remove as many source counts as possible from the raw data,
we advocate an aggressive approach to identifying the pixels to
be cleansed: the threshold significance should be set high, e.g.~to
$S_o = 10^{-2}$ (although never higher than 0.05, which corresponds to
the oft-used 95\% rejection level of statistics).
Regarding step (2), while masking is used by Damiani et al.~(1997a) in
their two-iteration approach to source detection, it would preclude us
from using FFTs to calculate $C(D_n)$.
Thus we replace the data in cleansed
pixels with the inferred background amplitude.

There are no rigorous quantitative rules governing how one should specify
the number of iterations, as that can depend on the
crowdedness of the field, the source distribution,
the source strengths, and the wavelet scale size.\footnote{For
a typical, uncrowded {\it Chandra} field, two iterations (i.e.~one
round of source count cleansing) are usually sufficient, because 
the high resolution of {\it Chandra} reduces source crowding relative
to that observed in, e.g., {\it ROSAT} data.  However, one should
always verify that this is the case with one's specific image!
}
We do note that iterative cleansing will cease 
if the background map does not change from one iteration to the next,
i.e., if no new pixels are marked for cleansing.

\subsubsection{Treating the Effect of Exposure Variations on Source Detection}

\label{sect:wexp}

In \S\ref{sect:thresh}, we describe how we use
probability sampling distributions $p(C \vert B_{i,j})$
to identify sources.  These distributions are derived assuming
a spatially constant exposure.  If, for
instance, the exposure map exhibits localized high-order variations,
then the list of detected sources may contain a mixture
of astrophysically interesting sources and ``instrumental sources''
aligned near support rib
shadows, near the edge of the FOV, at the location of hot pixels, etc.
(See, {\it e.g.}, Figure 1 of Damiani et al.~1997b.)
Thus, an efficacious source detection algorithm should include
additional calculations that act to decrease the detectability
of instrumental sources, while leaving the detection efficiency
of astrophysical sources unchanged.\footnote{
The distributions $p(C \vert B_{i,j})$ are also derived assuming a
spatially constant background map, from which simulated data are sampled.
Thus an efficacious source detection algorithm should also include
calculations that mitigate the effect of background variations
caused by, {\it e.g.}, X-ray shadows.  The current algorithm does
not take such variations into account; anecdotal evidence 
(e.g.~in {\S}\ref{sect:ver3}) indicates that they have little effect,
possibly because they are far less ``sharp" than variations induced
by support rib shadows, etc.
Note that if the background is known {\it a priori}, one can in principle 
remove the effect of high-order background variations on correlation
values using
a transformation similar to the one described below.}$^,$\footnote{
Note that exposure corrections are not
mandatory--for instance, the user may choose to have no corrections 
made if the analysis goal is scale-by-scale characterization of sources in 
correlation space.  See the caveats below.}

One possibility is to 
construct new sampling distributions $p(C \vert B_{i,j},E)$ for each
observation, taking into account all the exposure variations that can
appear within the wavelet support; however, this is not computationally
practical.
Instead, we estimate the systematic effect that exposure variations 
have on the correlation coefficients.  Assuming the null hypothesis, 
we may write the observed correlation coefficient as
\begin{equation}
C_{i,j}~=~<W{\ast}D>_{i,j}~=~<W{\ast}B>_{i,j}~+~{\Delta}C_{i,j} \,,
\end{equation}
where $B_{i,j}$ is the estimated (noise-free) background intensity
and ${\Delta}C_{i,j}$ is the noise 
(and possibly source count) contribution to $C_{i,j}$.
We ignore the latter term (see the caveats 
below) and rewrite the former so that
its dependence on exposure variations is explicit:
\begin{eqnarray}
<W{\ast}B>_{i,j}~&=&~<W{\ast}EB_{\rm norm}>_{i,j} \nonumber \\
&=&~\sum_{i'} \sum_{j'} W_{i-i',j-j'} E_{i',j'} B_{{\rm norm},i',j'} \nonumber \\
&=&~\sum_{i'} \sum_{j'} W_{i-i',j-j'} \left( E_{i,j} B_{{\rm norm},i',j'} - (E_{i,j}-E_{i',j'}) B_{{\rm norm},i',j'}  \right) \nonumber \\
&=&~E_{i,j} <W{\ast}B_{\rm norm}>_{i,j} - <W{\ast}({\delta}E B_{\rm norm})>_{i,j} \,. \label{eqn:expcor_deriv}
\end{eqnarray}
The quantity ${\delta}E_{i,j;i',j'}$ encapsulates exposure variability within 
the wavelet support, and thus the last term in eq.~(\ref{eqn:expcor_deriv})
encapsulates the effect of exposure variability upon $C_{i,j}$; subtracting
this term from $C_{i,j}$ yields an ``exposure-corrected" quantity that
contains only information of astrophysical value:
\begin{eqnarray}
C_{{\rm cor},i,j}~&=&~C_{i,j} - <W{\ast}({\delta}E B_{\rm norm})>_{i,j} \nonumber \\
&=&~C_{i,j} - <W{\ast}EB_{\rm norm}>_{i,j} + E_{i,j} <W{\ast}B_{\rm norm}>_{i,j} \label{eqn:fullcor} \,.
\end{eqnarray}
It is this quantity that is compared with the
distribution $p(C \vert B_{i,j})$ to determine whether $(i,j)$ is
a source pixel.


If $B_{\rm norm}$ is constant (or linear) within the wavelet support,
then eq.~(\ref{eqn:fullcor}) reduces to
\begin{equation}
C_{{\rm cor},i,j}^{\rm approx}~=~C_{i,j}~-~B_{{\rm norm},i,j}<W{\ast}E>_{i,j} .
\label{eqn:fastcor}
\end{equation}
Because $<W{\ast}E>$ is computed only once, we dub this the
``fast" exposure correction, as opposed to the ``full" exposure correction of 
eq.~(\ref{eqn:fullcor}).  One should not use the ``fast" correction if
non-linear structures 
(caused, e.g., by X-ray shadows) are apparent in the background map.

\begin{figure}
\begin{center}
\includegraphics[angle=-90,scale=0.75]{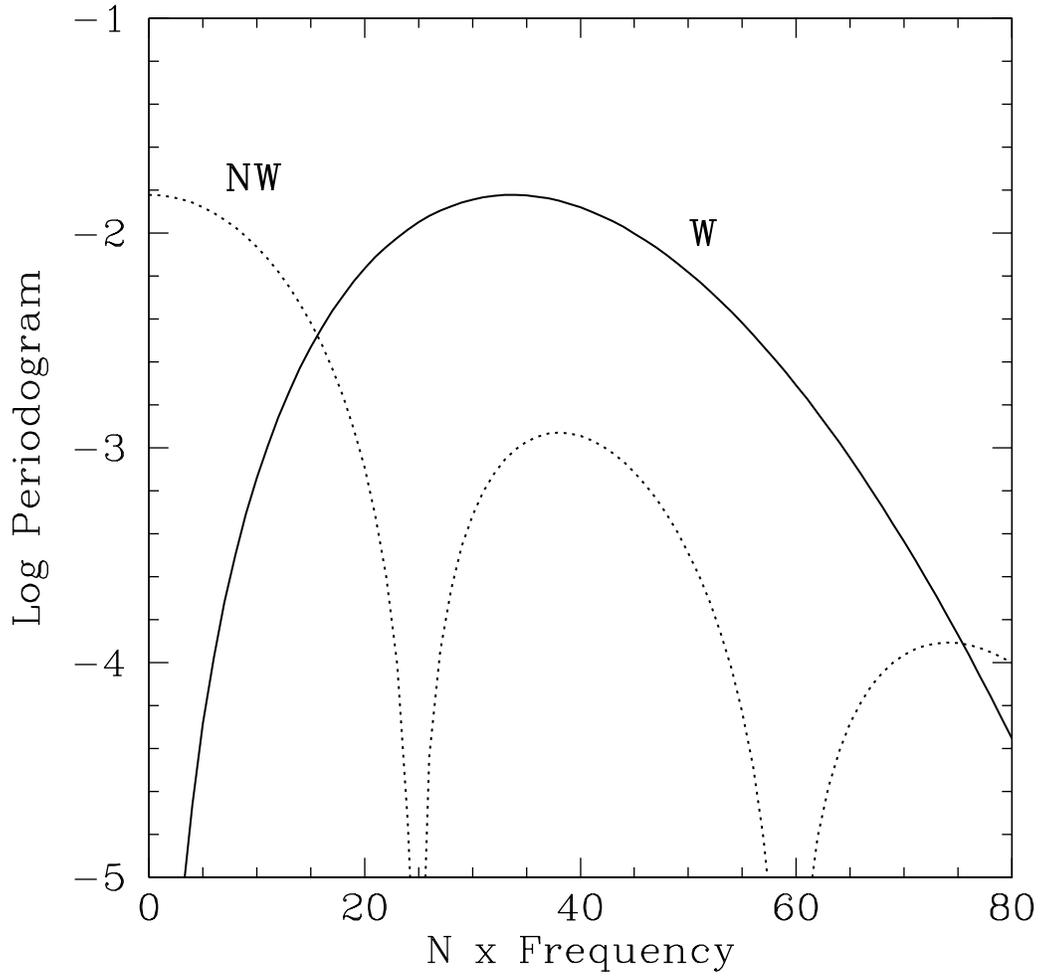}
\end{center}
\caption{
Sample power spectra of the Mexican Hat wavelet function ($W$) and
of the negative annulus of the Mexican Hat wavelet function ($NW$).
$N$ is the number of pixels in the (padded) image.
The $NW$ has much larger response to low-frequency components, but this
signal would be suppressed upon correlation with $W$ (see {\S}\ref{sect:wexp}).
}
\label{fig:pow}
\end{figure}

One should keep the following caveats in mind:
\begin{enumerate}
\item Strictly speaking,
$C_{{\rm cor},i,j}$ still cannot be directly compared with
the probability sampling distribution $p(C \vert B_{i,j})$ because
the noise term ${\Delta}C_{i,j}$ is itself uncorrected.  
If we concentrate on the issue of false positives ({\it i.e.}~assume
that there is no source count contribution to ${\Delta}C_{i,j}$), 
the important question is: 
is the asymptotic width of the distribution from which 
${\Delta}C_{{\rm cor},i,j}$ is sampled {\it smaller}
than the width of the distribution from which ${\Delta}C_{i,j}$ is
sampled?  If so, then the rate of false detections will still be greater
than expected.
This is a problem if and only if for a given pixel, 
$E_{i,j}$ is smaller than the average exposure $E_{\rm ave}$ 
over the wavelet support $(i,j)$, i.e.~this
is only a problem within troughs or beyond the edge of the FOV.
To see this, harken back to the simple example of {\S}\ref{sect:simp}:
what would happen to the width of $p(C|B_{i,j})$ if we were to reduce the
exposure?  Fewer counts would be detected, so $B_{i,j}$ would decrease,
and the width of the noise distribution, which is $\propto \sqrt{B_{i,j}}$,
would also decrease.  Thus we suggest that one should carefully
scrutinize all sources detected in 
low-exposure regions ($\simless 0.2 E_{\rm max}$).

\item Note the distinction between the correlation
maps $C$ and $C_{\rm cor}$, especially if the analysis goal is not
just source detection, but also image 
decomposition (the scale-by-scale characterization of sources 
in correlation space).\footnote{
Note that source characterization in {\tt WAVDETECT} is done using the raw data
themselves and {\it not} using correlation coefficients.
Aside from source detection, the only other place where the exposure-corrected
correlation map is used in {\tt WAVDETECT} is in the creation of the
noise-free, exposure-corrected 
image of detected sources ({\S}\ref{sect:noise}).  Thus this
caveat is only an issue if the user desires to analyze correlation maps
{\it outside} of {\tt WAVDETECT}.}
The quantity $C_{\rm cor}$ 
represents a {\it pixel-by-pixel} wavelet filtering not of the raw
data $D$, but of the
quantity $D - EB_{\rm norm} + E_{i,j}B_{\rm norm}$ (eq.~\ref{eqn:fullcor}).
Because $B_{\rm norm}$ may be estimated using the $NW$, which is
most sensitive to low-frequency components of the data (see
Figure \ref{fig:pow}), these modified
data may be ``contaminated" with low-frequency information (although
wavelet filtering [eq.~\ref{eqn:fullcor}]
mitigates the effect of the contamination).

\item We note that while systematic overestimates of $B_{\rm norm}$ 
(caused for reasons discussed in {\S}\ref{sect:back})
adversely affect the computation of $C_{\rm cor}$, they will not lead to an 
increased number of false detections.  This is because the only situation
where such a systematic overestimate affects {\it non-source} pixels in
the {\it final}, refined background image is when the background is computed
in regions of extended diffuse emission.
In this situation, the
algorithm treats the sum of the real background and the diffuse emission
as the ``background," so the rate of false detections (which
is independent of background amplitude) will be unchanged.

\end{enumerate}

\subsubsection{Variance Estimation}

\label{sect:werror}

We estimate the variances of $B_{i,j}$ (if one does not provide a background
map), and of $C_{i,j}$ (or $C_{{\rm cor},i,j}$ or 
$C_{{\rm cor},i,j}^{\rm approx}$), using the standard formula 
(Eadie et al.~1971, p. 23)
\begin{eqnarray}
V[Y]~&=&~V[\sum_i \sum_j a_{i,j} X_{i,j}] \nonumber \\
&=&~\sum_i \sum_j a_{i,j}^2 V[X_{i,j}] + 2 \sum_i \sum_{i' > i} \sum_j \sum_{j' > j} a_{i,j} a_{i',j'} {\rm cov}[X_{i,j},X_{i'j'}] \label{eqn:covar} 
\end{eqnarray}
where $Y$ is quantity of interest and $X_{i,j}$ are random variables
(either $D_{i,j}$ or functions of $D_{i,j}$).  For instance,
\begin{eqnarray}
V[C_{i,j}]~&=&~V[\sum_{i'} \sum_{j'} W_{i-i',j-j'}D_{i',j'}] \nonumber \\
&=&~\sum_{i'} \sum_{j'} W_{i-i',j-j'}^2 V[D_{i',j'}] \nonumber \\
&=&~\sum_{i'} \sum_{j'} W_{i-i',j-j'}^2 D_{i',j'}\label{eqn:corerr2} \\
&=&~<W^2{\star}D>_{i,j} \,. \label{eqn:corerr3}\nonumber
\end{eqnarray}
Note that we make two assumptions when deriving this formula:
(1) the datum $D_{i',j'}$ is sampled from
a Poisson distribution with variance $D_{i',j'}$; and (2) each pixel's
raw datum is independently sampled (so covariance terms do not
contribute to $V[C_{i,j}]$).

We list variance formulae related to source detection in Table \ref{tab:sderr}.
The reader should keep in mind two important caveats about them:
\begin{enumerate}
\item These formulae ignore the contribution of
the covariance terms, which are non-zero
for $V[C_{\rm cor}]$, $V[C_{\rm cor}^{\rm approx}]$, and 
$V[B]$ if the data are iteratively cleansed, i.e.~if
$B$ is a not just a function of the raw data $D$ only.
We ignore these terms because even the simplest covariance computation,
that for a two-iteration background map (see Appendix \ref{sect:covar}),
has a staggeringly high computational cost: we find that
the CPU time needed to compute the variance increases by a factor
$\sim {\cal O}(d_xd_y\sigma_x^2\sigma_y^2)$, where $d_x$ and $d_y$ are the
x- and y-axis lengths in pixels, respectively.
Also, additional
arrays containing information needed to compute the covariance terms must
be kept in memory, so there is a resource cost as well.
We find that including covariance terms increases the variance by
a median value of $\approx$ 7\%, and at most by
only $\approx$ 30\% adjacent to strong sources, although this is a
source-strength- and source-geometry-dependent result that obviously
cannot be blindly applied to all fields.  Ultimately it is up to the user to
judge whether adding the computation of covariance to our base algorithm
is worthwhile.
\item When computing $V[B_{i,j}]$ for the final background map,
we make the simplifying assumption that the variance of a
cleansed datum is equal to the cleansed datum itself; for instance,
for a two-iteration background map,
\begin{eqnarray}
V[B_{i,j}]~&=&~\sum_{i'}\sum_{j'} a_{i,i',j,j'}^2 V[D_{2,i',j'}] \nonumber \\
&=&~\sum_{i'}\sum_{j'} a_{i,i',j,j'}^2 D_{2,i',j'} \label{eqn:bvar} \,,
\end{eqnarray}
where
\begin{equation}
a_{i,i',j,j'} = E_{i,j} \frac{NW_{i-i',j-j'}}{<NW{\star}E>_{i,j}} \,.
\end{equation}
We make this assumption because if the data
are a mixture of raw data and background
estimates, the variance estimates become
increasingly complicated: for one iteration,
\begin{eqnarray}
V[D_{i,j}]~&=&~D_{i,j} \,, \nonumber
\end{eqnarray}
for two iterations,
\begin{eqnarray}
V[D_{2,i,j}]~&=&~\cases{ D_{i,j} & {\rm uncleansed pixel} \cr
\sum_{i'}\sum_{j'} a_{i,i',j,j'}^2 D_{i',j'} & {\rm cleansed pixel}} \label{eqn:twoiter}
\end{eqnarray}
etc.
\end{enumerate}

\subsection{Source Characterization}

\label{sect:wrecon}

\begin{figure}
\begin{center}
\includegraphics[scale=0.75]{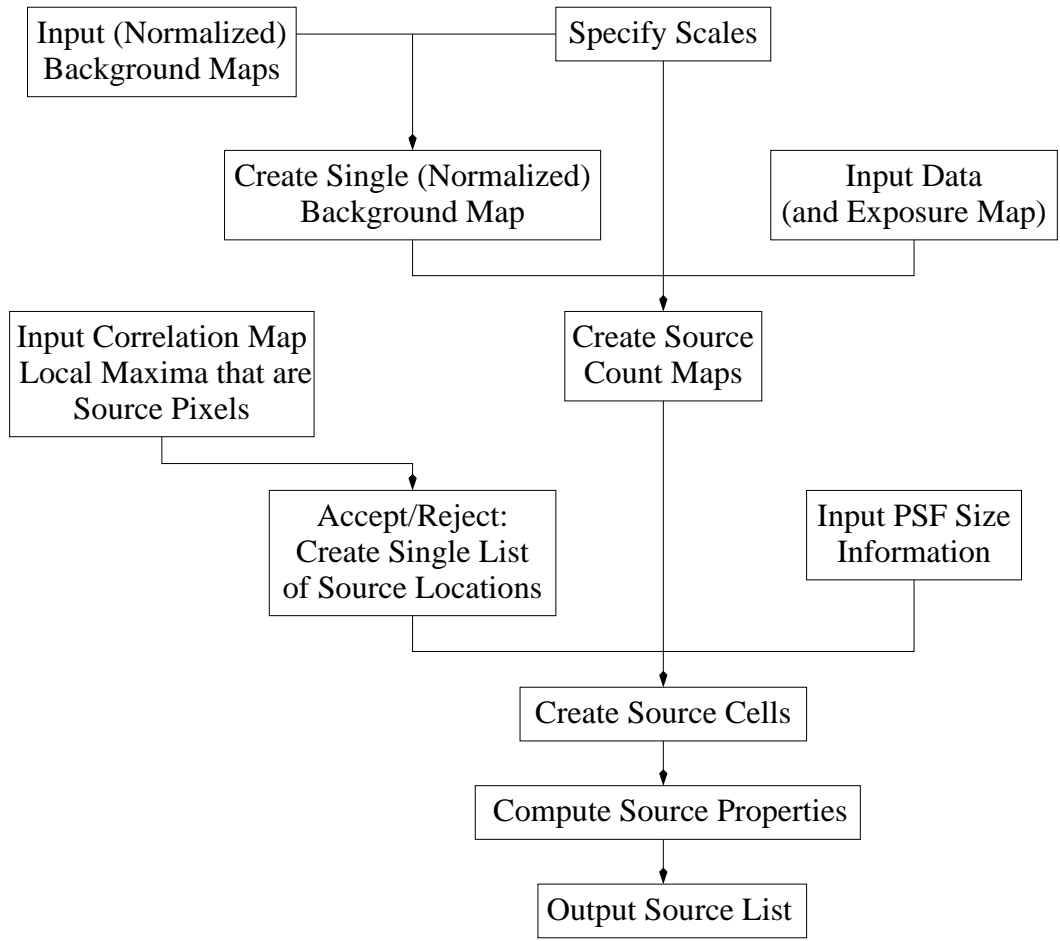}
\end{center}
\caption{
Flow chart illustrating the source list generation algorithm 
described in {\S}\ref{sect:wrecon}.  Parantheses indicate optional
input.
}
\label{fig:wrflow}
\end{figure}

Once we have identified putative source 
pixels at each of a number of wavelet scale size pairs
$(\sigma_x,\sigma_y)$, our next objective is {\it source characterization},
wherein we combine information derived at each scale pair to
generate a final source list and to estimate source properties
(see Figure \ref{fig:wrflow}).
Unlike source detection, source characterization algorithms can be
arbitrarily complex, depending, for instance, upon whether one wishes 
to use detailed PSF information.
Our method is particularly simple, in that we use only
the characteristic PSF size at a given pixel, $r_{\rm PSF,i,j}$.
This size may be associated with, e.g., 50\% encircled energy;
we find that smaller values, such as 39.3\% (which corresponds to the
integral of a symmetric normalized two-dimensional Gaussian to radius 
$\sigma_G$), work better than large values.
While using the detailed PSF shape may allow for more accurate estimates of
source properties, the simplicity of our scheme makes it
more immediately applicable to images from virtually any
counts detector (including those for which calibration is on-going or
is for other reasons incomplete).

\subsubsection{Corrected Background Estimate}

\label{sect:nbkg}

In {\S}\ref{sect:back}, we describe how we calculate a 
scale-dependent normalized local background estimate $B_{\rm norm}$,
by assuming that there are no source counts in the negative annulus, $NW$.
Source detection itself is not markedly affected if this assumption
is violated and the background overestimated, 
for reasons given in {\S}\ref{sect:back}, but an overestimated
background will adversely affect the computation of source
properties.
We create a new, corrected, background estimate by
combining information across scales (denoted with
a subscript $k$), noting that the assumption that
there no source counts in $NW$ is always violated around sources if
either $\sigma_{x,k}$ or $\sigma_{y,k}$ is less than the source size:
\begin{equation}
B'_{{\rm norm},i,j}~=~\frac{\sum_{k=1}^N \epsilon_{i,j,k} \sigma_{x,k} \sigma_{y,k} B_{{\rm norm},i,j,k}}{\sum_{k=1}^N \epsilon_{i,j,k} \sigma_{x,k} \sigma_{y,k}}.
\label{eqn:bsum}
\end{equation}
$N$ is the number of scale pairs used, and
\begin{equation}
\epsilon_{i,j,k}~=~\cases{1&${\rm min}(\sqrt{2}\sigma_{x},\sqrt{2}\sigma_{y}) \geq mr_{{\rm PSF},i,j}$\cr
0&otherwise\cr}
\label{eqn:bsumm}
\end{equation}
$m$ is a multiplicative factor, set to one when estimating the properties
of point sources, and to larger values when extended sources are analyzed.
We use eq.~(\ref{eqn:covar}) 
to estimate the variance of $B'_{{\rm norm},i,j}$, with
covariance terms ignored:
\begin{equation}
\sum_{k=1}^N \left(\frac{\epsilon_{i,j,k}\sigma_{x,k}\sigma_{y,k}}{\sum_{k=1}^N \epsilon_{i,j,k}\sigma_{x,k}\sigma_{y,k}}\right)^2 V[B_{{\rm norm},i,j,k}] \,.
\label{eqn:corbvar}
\end{equation}
$V[B_{{\rm norm},i,j,k}]$ is estimated using the approximate equation
listed in Table \ref{tab:sderr}.

\subsubsection{Source Cells}

\label{sect:cell}

The next step in source characterization is to determine which pixels of
the original image $D$ are to be associated with each detected source.
We term contiguous pixels which are associated with a particular source
a {\it source cell}, and as we describe below in {\S}\ref{sect:sprop}, we
use the raw data in a source cell to determine a source's properties.
We need to create source cells because, as noted above, we do not use
PSF shape information in our algorithm, 
and because we want an algorithm that is applicable 
to both point and extended sources.  Source cells are not computed in
algorithms such as the sliding cell, where the integrated 
PSF volume and sum of (point) source counts for a given user-defined
cell can be used to estimate the total number of (point) source counts,
etc.

\begin{figure}
\begin{center}
\includegraphics{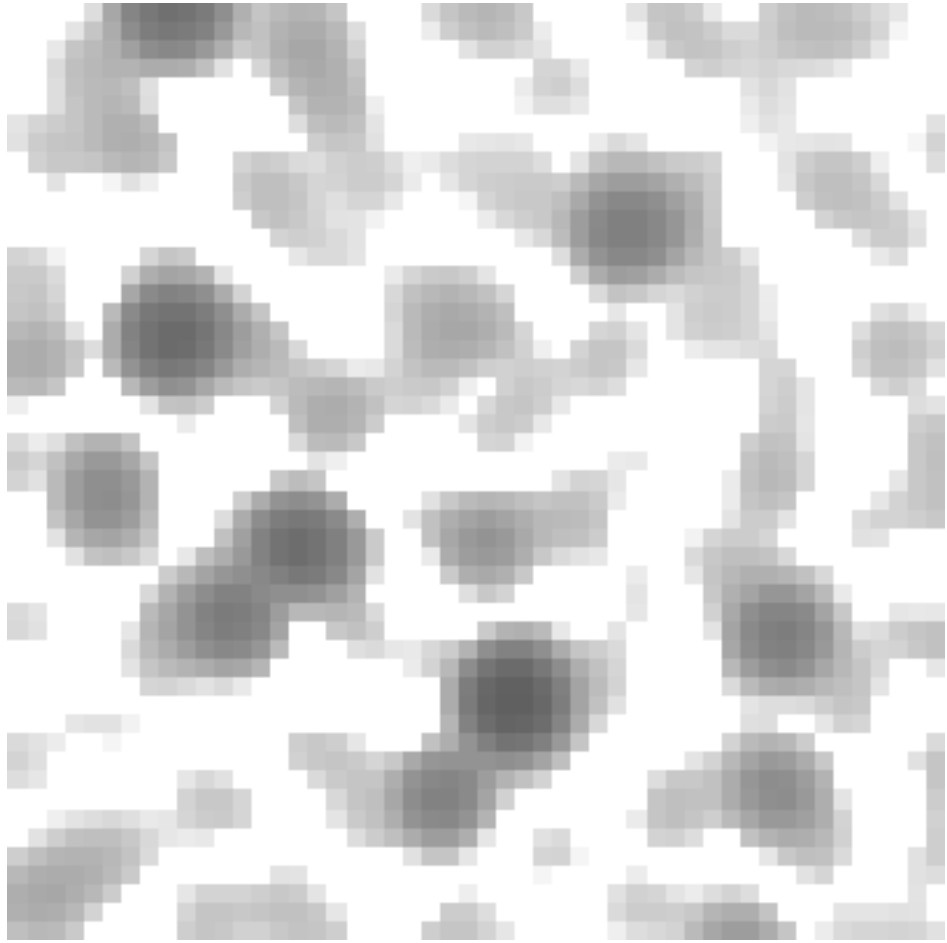}
\end{center}
\caption{
A sample source counts image ({\S}\ref{sect:cell}), created by first 
smoothing the data in a 50$\times$50 subfield of the 
{\it ROSAT} PSPC Pleiades Cluster image
with a $PW$ function of size $\sigma_x$ = $\sigma_y$ = 2 pixels,
then subtracting the estimated background.  Pixels associated
with the strongest local maxima comprise the source cells that are
used for source property estimation.  See Figures \ref{fig:scex1}
and \ref{fig:scex2}.
}
\label{fig:scounts}
\end{figure}

To create source cells, we first compute source count images,
smoothing the raw data with the positive kernel of the wavelet function,
$PW$, at user-specified scales,
and then subtracting the corrected background image, $B'$:
\begin{equation}
SC_{i,j,k}~=~{\rm max}\left(\frac{<PW{\star}D>_{i,j,k}}{<PW{\star}E>_{i,j,k}}~-~B'_{{\rm norm},i,j},0\right) .
\label{eqn:flux}
\end{equation}
(We denote these images $SC$ to avoid confusion with either the significance
$S$ or the correlation image $C$.)
We use $PW$ as the smoothing function
because it has the desirable properties of being localized,
and, for the particular case of a symmetric MH function, 
of mimicking the shape of a canonical Gaussian PSF.
In regions where there are no sources, source count image values
are either zero, or positive and nearly zero;
only in the vicinity of sources do the values deviate markedly from zero.
Thus the source count images appear to contain numerous ``islands" of
non-zero flux in a sea of zero values, with their relative size
increasing with size of the smoothing function $PW$ (see Figure 
\ref{fig:scounts}).  Each island
contains one or more peaks, and sub-islands may be defined using each
peak, with saddle points providing the boundaries between them.
(Sub-)islands observed in selected source count images
define the source cells.

To define a source cell, we must select a source counts image and
then determine to which (sub-)island the putative source belongs.
The selection proceeds as follows.
The location of a putative source in correlation-space is assumed to be
the location of the correlation maximum, $(i_C,j_C)$.
At this location, we compute the PSF size, in pixels,
$r_{{\rm PSF},i_C,j_C}$.  (This introduces a bias towards
point sources; we return to this point below.)
We then select the source count image 
with smoothing scale ``closest" to $r_{{\rm PSF},i_C,j_C}$
by minimizing
$\vert {\log}_2 \sigma_k - {\log}_2 r_{{\rm PSF},i_C,j_C} \vert$, where
$\sigma_k$ is defined for each scale pair:
\begin{eqnarray}
\sigma_k~&=&~\exp\left[\frac{\log(\sigma_x)+\log(\sigma_y)}{2}\right] \,. \nonumber
\end{eqnarray}
On the selected source count image, we examine pixel $(i_C,j_C)$:
the (sub-)island to which this pixel belongs defines the source cell.

\begin{figure}
\begin{center}
\includegraphics[scale=0.75]{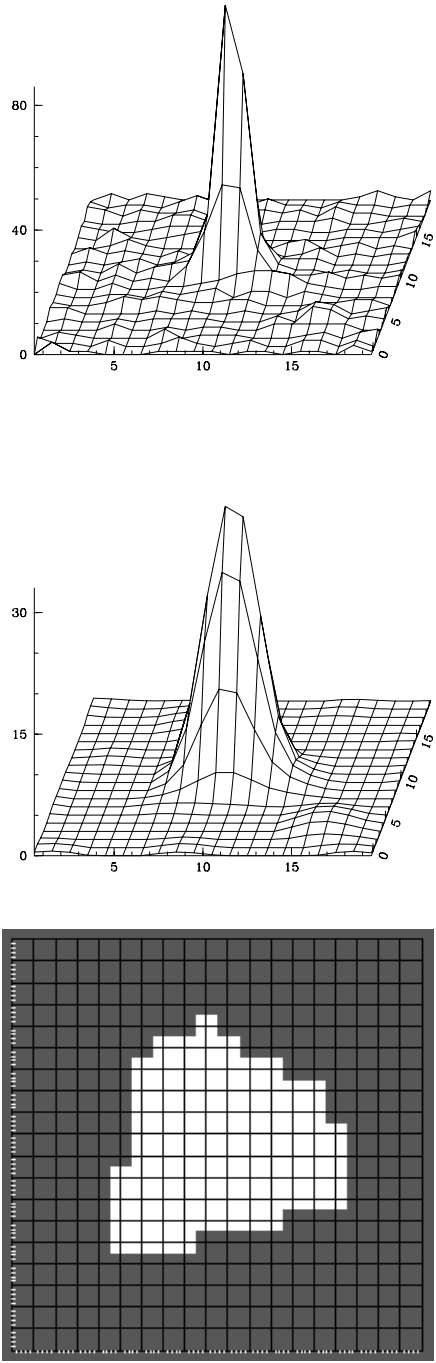}
\end{center}
\caption{
Top: counts data showing an isolated Pleiades Cluster source, 
as observed by the {\it ROSAT} PSPC.
Middle: source counts image data, created smoothing the counts data 
with a $PW$ function of size $\sigma_x$ = $\sigma_y$ = 2 pixels,
then subtracting the estimated background.
Bottom: the source cell defined using the source counts image data 
({\S}\ref{sect:cell}).
This cell, used in the estimation of source properties, 
contains nearly all, if not all, of the counts from this source.
}
\label{fig:scex1}
\end{figure}

\begin{figure}
\begin{center}
\includegraphics[scale=0.75]{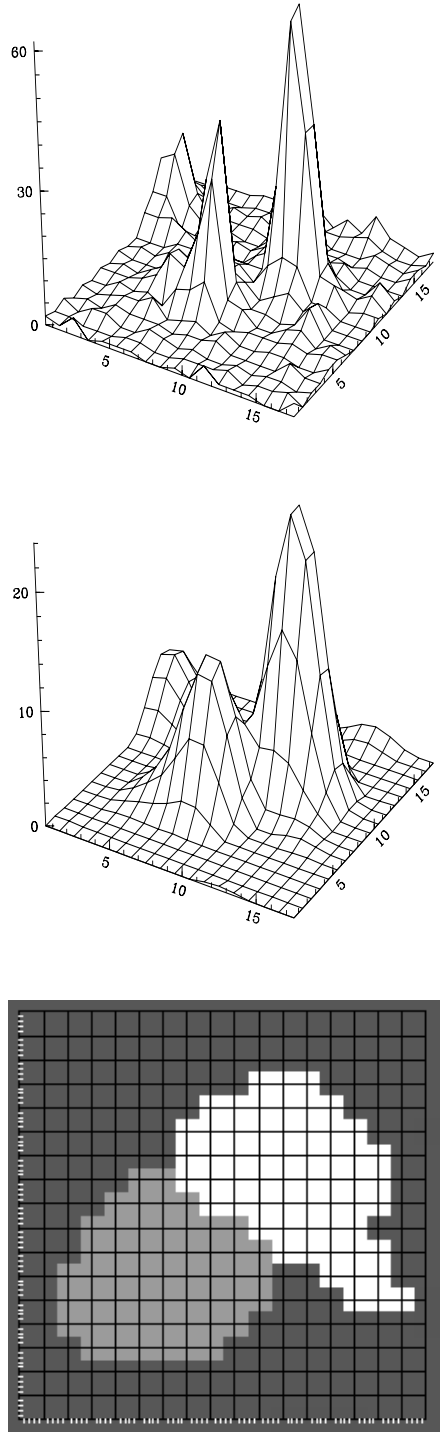}
\end{center}
\caption{
Top: counts data showing two nearly overlapping Pleiades Cluster sources, 
as observed by the {\it ROSAT} PSPC.
Middle: source counts image data, created smoothing the counts data 
with a $PW$ function of size $\sigma_x$ = $\sigma_y$ = 2 pixels,
then subtracting the estimated background.
Bottom: the source cells defined using the source counts image data.
The saddle point seen in the middle image defines the boundary
between the cells ({\S}\ref{sect:cell}).
}
\label{fig:scex2}
\end{figure}

A source cell defined in this manner has advantageous properties:
(1) if $\sigma_k \approx r_{{\rm PSF},i_C,j_C}$, then nearly
all isolated point source counts should lie within a source cell
(Figure \ref{fig:scex1}); (2)
exposure variations are taken into account via the use of
$<PW{\star}E>$ in eq.~(\ref{eqn:flux}), so that source cells are not
truncated near, e.g., support rib shadows; and (3), as noted above,
saddle points in the source count images provide natural boundaries between
sources in crowded fields (Figure \ref{fig:scex2}).

We note two situations where care must be exercised when
interpreting results.  First, 
the source cell for an extended source may be too small
if the smoothing scale is $\approx r_{{\rm PSF},i_C,j_C}$.
The steps one must take to deal with this situation will vary depending upon
analysis circumstances, but one possible step is to create 
only one source count image, with
$\sigma_k \gg r_{{\rm PSF},i_C,j_C}$, and to use this image to define
the cell, while being careful to note whether previously detected point
sources are located within it.  (See, e.g., {\S}\ref{sect:ver1}.)
Another situation for which care must be exercised is when
the PSF is bimodal or otherwise strangely behaved (such as the off-axis
{\it Chandra} PSF); two or more source cells could be created for one detected
source.  The necessary steps to deal with this situation depend upon
the details of the detector itself, and thus we will not discuss this
particular situation further here.

\subsubsection{Source Rejection}

Because the same source will generally be detected at multiple
scales, and to further decrease the possibility of finding false
sources, it is necessary to reject sources from the lists 
of correlation maxima generated
at each scale.
A maximum observed at $(i_C,j_C)$,
for the scale pair $(\sigma_{x,k},\sigma_{y,k})$, is
rejected from further consideration if any of the following conditions
are met:
(1) $(i_C,j_C)$ lies in a previously defined source cell;
(2) $SC_{i_C,j_C,k} =$ 0;
(3) the ellipse defined by
\begin{eqnarray}
\frac{(x-i_C)^2}{2\sigma_{x,k}^2}~+~\frac{(y-j_C)^2}{2\sigma_{y,k}^2}~=~1 \nonumber
\end{eqnarray}
contains one or more previously defined
sources detected at smaller scales
(this can occur when $\sigma_{x,k}$ or $\sigma_{y,k}$ is $\simgreat
r_{\rm PSF}$, since previously identified sources will eventually merge
if the field is crowded, creating ``new" sources at new locations);
and (4)
if, after the source cell is defined for a particular scale, it
is found not to contain any correlation maxima {\it at that scale}.
This last check is aimed at rejecting small-scale Poisson fluctuations
that may be observed in the background data.

\subsubsection{Source Properties}

\label{sect:sprop}

We present the formulae we use to estimate source properties
and their variances in Tables \ref{tab:prop} and \ref{tab:error}
respectively.
The summations performed when making these estimates are carried out over
the pixels within the source cell.
We use the raw counts data, $D_{i,j}$, as a weighting
function, instead of the source fluence
$D_{i,j} - E_{i,j}B'_{{\rm norm},i,j}$,
because the use of the latter can greatly complicate the
estimation of variances.
Using the data rather than the source fluence will lead to similar
estimates when the background amplitude is small relative to source
amplitude.

\subsubsection{Noise-Free Source Image}

\label{sect:noise}

We can use the information present in the correlation images and the 
source cell image to create a ``noise-free" rendering of the observed
source data, $SD$, with the effect of exposure variations removed:
\begin{equation}
SD_{i,j}~=~\sum_{k=1}^N \frac{C_{{\rm cor},i,j,k}}{\sigma_{x,k}\sigma_{y,k}} \nu_{i,j,k} .
\end{equation}
Dividing the correlation value
by $\sigma_{x,k}$ and $\sigma_{y,k}$ restores
the normalization contained in eq.~(\ref{eqn:wave}), allowing
a scale-by-scale summation.
The quantity $\nu_{i,j,k}$ = 1 if
(1) $C_{{\rm cor},i,j,k} >$ 0, 
(2) the local maximum corresponding to $(i,j)$ has been 
identified as a source pixel, 
and (3) the associated local correlation maximum is contained within a source
cell
(the second condition ensures that random
maxima which are not associated with a source but which 
happen to lie within a source cell are not included in the source image;
the last condition ensures that rejected sources
are not included);
otherwise, $\nu_{i,j,k}$ = 0.


\section{Verification}

\label{sect:ver}

To verify its source detection and characterization capabilities, we apply our
algorithm to: (1) 
1 and 10 ksec observations by an idealized detector with a spatially
invariant PSF;
(2) a 32-ksec {\it ROSAT} PSPC observation of the
Pleiades Cluster; and
(3) a simulated 
30-ksec {\it Chandra} ACIS-I observation of the Lockman Hole region.
These tests allow us to demonstrate that 
our algorithm can efficiently detect and accurately describe 
well-sampled sources in an uncrowded field, and can
effectively analyze crowded fields, even in the low-background
limit of the {\it Chandra} detectors.

\subsection{Idealized Detector with Spatially Invariant PSF}

\label{sect:ver1}

\begin{figure}
\begin{center}
\includegraphics[scale=0.75]{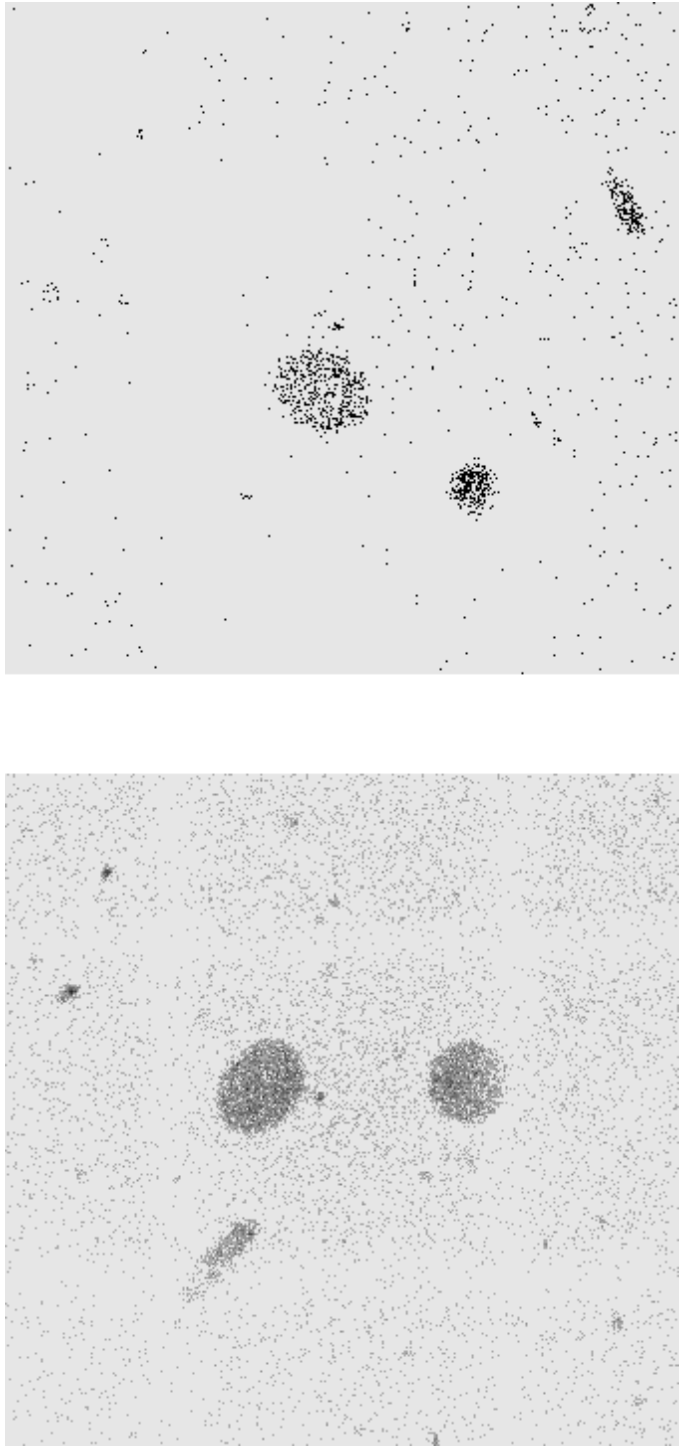}
\end{center}
\caption{
Top: a 512$\times$512 pixel image (Image A) showing a 1-ksec 
observation by an 
idealized detector with effective area 1000 cm$^2$, a spatially 
invariant Gaussian
PSF of width $\sigma_{\rm PSF}$ = 2.56 pixels, and an exposure map
similar to that of the {\it Einstein} IPC ({\S}\ref{sect:ver1}).  
Within this image were
placed 42 point and 4 extended sources.  The background is assumed to
be locally variable, with amplitude $\sim$ 10$^{-5}$ ct sec$^{-1}$
pix$^{-1}$.  
Bottom: same as top, except the observation time is 10 ksec (Image B).
}
\label{fig:simdata}
\end{figure}

We first demonstrate that our algorithm
can efficiently detect, and accurately describe,
well-sampled sources in an uncrowded field.
We apply it to two 512$\times$512 images, hereafter Images A and B,
that represent 1 and 10 ksec observations by
an idealized detector with an effective area 1000 cm$^2$ and a
spatially invariant
Gaussian PSF of width $\sigma_{\rm PSF} =$ 2.56 pixels
(Figure \ref{fig:simdata}).
The exposure map for this detector is similar to
that of the {\it Einstein} IPC.
Within each image we randomly place 42 point sources, and 4 extended
sources with elliptical shape.  The fluxes of the point sources were
sampled from a ${\log}N-{\log}S$ distribution with slope $-1.5$, above
10$^{-14}$ erg cm$^{-2}$ sec$^{-1}$.
We also simulate a locally variable background 
(amplitude $\sim$ 10$^{-5}$ ct sec$^{-1}$ pix$^{-1}$)
by setting the background amplitude at five reference points,
performing minimum-curvature-surface interpolation,
and sampling background data in each pixel.

\begin{figure}
\begin{center}
\includegraphics[angle=-90,scale=0.60]{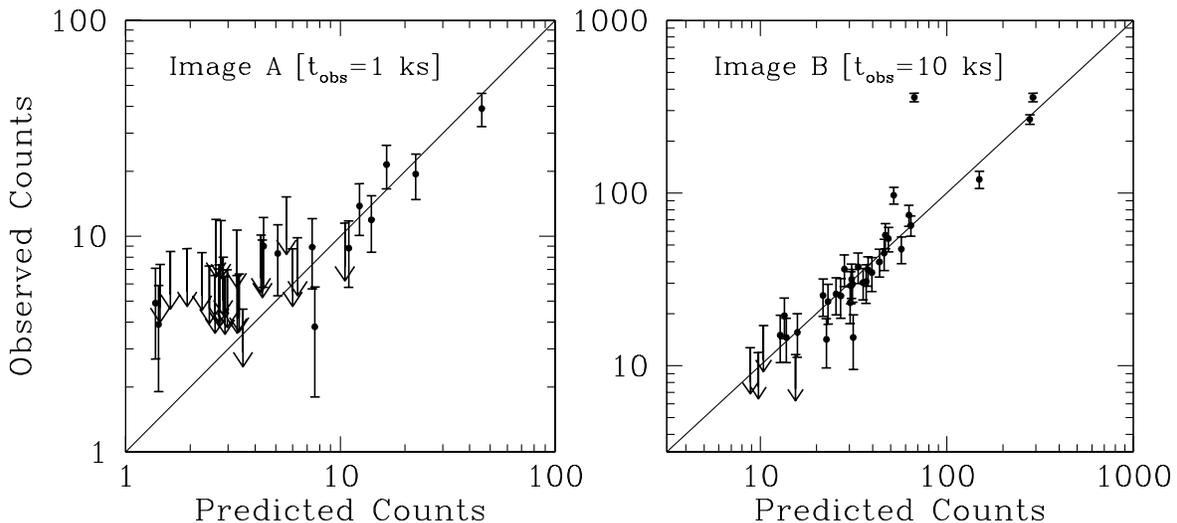}
\end{center}
\caption{
Top: comparison of observed counts, with estimated 1$\sigma$ errors, 
and upper limits for undetected sources, with
predicted counts for the point sources of Image A.
Upper limits are defined using the source detection threshold values
at the correlation maxima nearest the location of the undetected 
sources, and are computed using eq.~(\ref{eqn:cmax}), with
$\sigma_x = \sigma_y = \sqrt{3}\sigma_G =$ 4.43 pixels.
The source exposure time (see Table \ref{tab:prop}),
not the total observation time,
is used to compute the predicted counts.
Bottom: same as top, but for Image B.  The two farthest outliers on this figure
correspond to observed ``sources" which are actually composed of two
sources each.
}
\label{fig:counts}
\end{figure}

We analyze the images assuming the input parameters listed for
Test 1 in Table \ref{tab:numdet}.
In Figure \ref{fig:counts}, we plot the source counts
for detected sources, and the upper limits for undetected sources,
against the number of predicted counts.
Upper limits are defined using the source detection threshold values
at the correlation maxima nearest the location of the undetected 
sources, and are computed using eq.~(\ref{eqn:cmax}), with
$\sigma_x = \sigma_y = \sqrt{3}\sigma_G =$ 4.43 pixels.
We conclude that our algorithm efficiently detects and describes point
sources with $\simgreat$ 10 counts.
This is not an absolute quantity: the minimum number of counts 
needed for source detection varies as a function of the background
amplitude and source size (see {\S}\ref{sect:simp}).
Hence while we may conclude that a 1 ksec observation
is sufficient for the detection by {\it Chandra} of nearly on-axis
point sources with fluxes $\simgreat$ 10$^{-14}$ erg cm$^{-2}$ sec$^{-1}$,
since it has a similar effective area as, and lower expected
background count-rates than, our idealized detector,
it may not be sufficient far off-axis,
or for detectors that have higher rates of background accumulation.

\begin{figure}
\begin{center}
\includegraphics[scale=0.75]{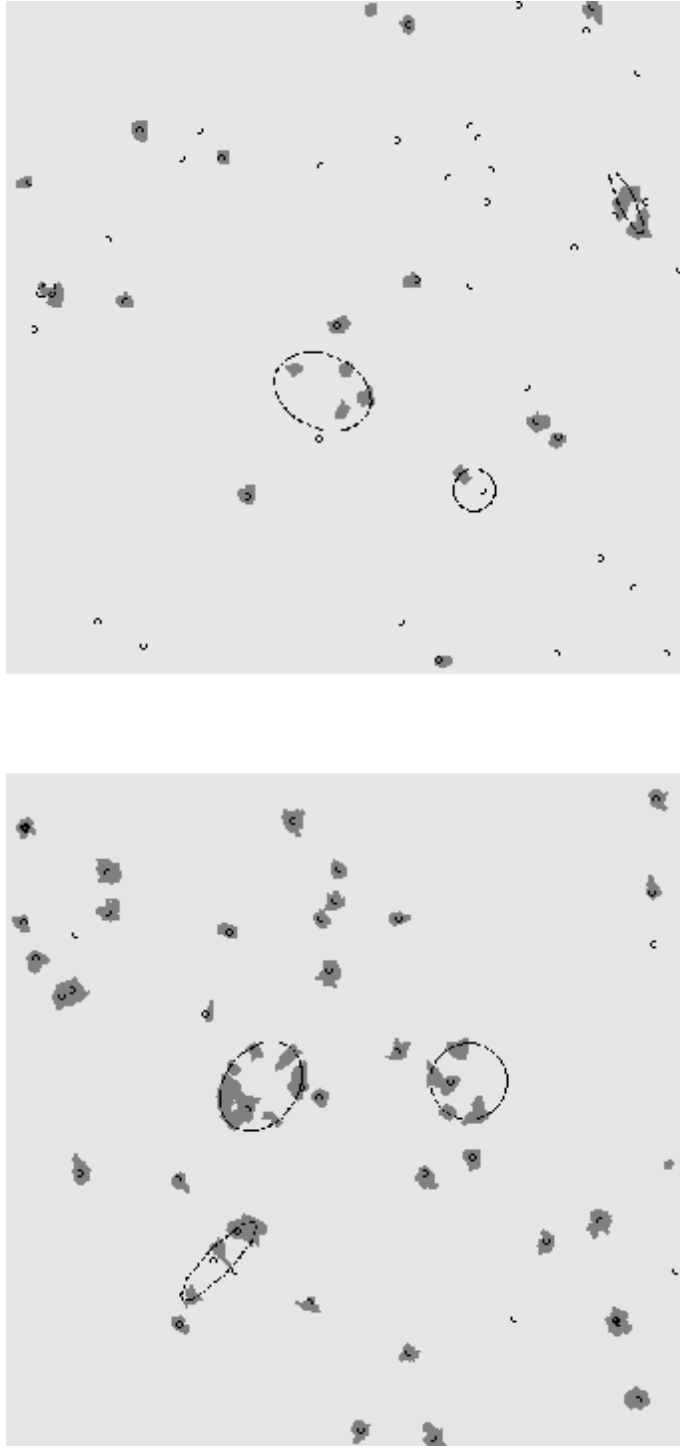}
\end{center}
\caption{
Top: source cell image generated during the analysis of Image A,
for wavelet scale size 2$\sqrt{2}$ pixels.
The overlaid small circles and larger ellipses represent respectively 
the 42 point and 4 extended sources randomly placed in the image.
Bottom: same as top, but for Image B.
}
\label{fig:cell}
\end{figure}

In Figure \ref{fig:cell},
we show the cells for the detected sources, along with the
input source locations.
We use a wavelet scale of 2$\sqrt{2}$ pixels
to create the source counts image that is in turn
used to delineate the source cells.
This scale is the closest ``standard" scale to the assumed PSF size
(if we assume scale sizes separated by factors of $\sqrt{2}$ rather
than 2 for greater source detection efficiency).
In both images, our algorithm detects one false source, which is consistent
with the assumption of $S_o = 10^{-6}$ for a 512$\times$512 image.

\begin{figure}
\begin{center}
\includegraphics[scale=0.70]{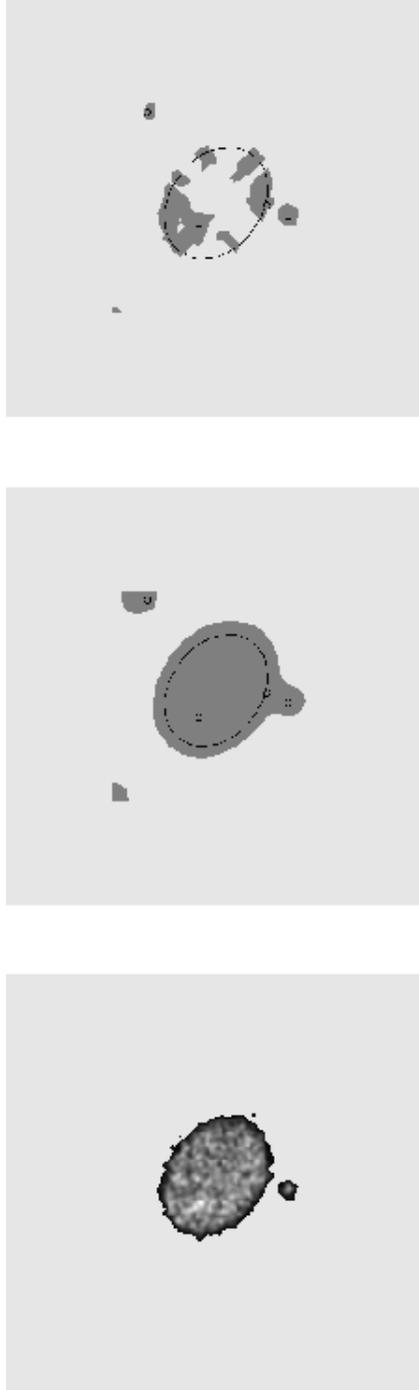}
\end{center}
\caption{
Top: subset of the source-cell image shown at the bottom of
Figure \ref{fig:cell}, showing 
the cells generated in the vicinity of the largest extended source in
Image B.
Middle: same as top, except that now the source cells are defined using a
source counts image with smoothing size 8$\sqrt{2}$ pixels instead of
2$\sqrt{2}$ pixels.
Bottom: the normalized count-rate image for the extended source (and three
point sources), generated
by creating a source counts image with smoothing size 2 pixels, then
filtering out all pixels not contained in the source cell shown immediately
above.  If
this source were a galaxy cluster, the analyst could proceed to fit
to these data (respecting the caveats discussed in {\S}\ref{sect:ver1}).
}
\label{fig:ext}
\end{figure}

Creating source cells by using information derived at a wavelet scale
close to the PSF size
is not optimal when one wishes to analyze and describe extended sources,
as discussed in {\S}\ref{sect:cell}, and indeed by examining
Figure \ref{fig:cell} we can see that the extended source cells are
undersized.
There are many ways by which an analyst may wish to treat extended 
sources; here, we show how one could derive an image showing the
(normalized) number of counts per pixel within the
extended source.  We use as our example 
the largest extended object in Image B.
First, we must expand the source cell so that it just encloses the extended
source.  To do this, we increase the minimum
scale size at which a source counts image 
is to be computed (in this example, from 2$\sqrt{2}$ pixels to 8$\sqrt{2}$ 
pixels; see Figure \ref{fig:ext}a-b).  (Note that 
the parameter $m$ in eq.~\ref{eqn:bsumm} must also be increased so that
the background is not overestimated within the source.)
Second, we would use the new source cell as a spatial filter,
applying it to the original source counts image created at the PSF scale
(e.g.~Figure \ref{fig:ext}c), or to the data, etc.

(If one outputs source count image data, one can then, 
in principle, fit directly to them.
For instance, the image may be of a galaxy cluster, and one may
wish to assess the detectability of its constituent
galaxies.  However, care must be exercised since the data
in contiguous pixels are not independent.  Taking into account the
width of the $PW$ used to smooth the raw image data, we can state that
data $> \sqrt{2}\sigma$ pixels apart are independent.  Thus simple statistical
fitting can be done to a sparse grid of data.  This process of fitting
would be essentially equivalent to the 
``decimation" method described by Lazzati et al.~1998, except that they fit to
correlation image data.)

\subsection{ROSAT PSPC: The Pleiades Cluster}

\label{sect:ver3}

Next, we demonstrate that our algorithm outperforms the sliding cell in
efficiently detecting sources in a crowded field, by applying it
to the deep (32 ksec) {\it ROSAT} PSPC observation
of the core of the Pleiades Cluster (RP200068, cf.~Micela et al.~1996).
These data were obtained in two segments separated by
roughly six months, and the slight boresight offset between the
two segments has been corrected using a method 
described by Micela et al.
The data were also filtered to exclude times of high background
contamination, and to exclude 
pulse-height-invariant (PI) channels at both the
low-energy (PI$<$20, to avoid the so-called ``ghost image''
problem; Nousek \& Lesser 1993) and high-energy (PI$>$201, where no
instrument map is available to determine exposure variations) ends
of the spectral response.  We have computed exposure maps taking
into account these changes using software developed by Snowden \&
Kuntz (1998).

In Table \ref{tab:numdet}, we show how the number of detected sources varies
as a function of the number of iterations, the spacing of scales,
the exposure correction method, and the source detection significance.
Comparing Tests 1 through 6, for which $S_o$ is constant,
we find that the number of detected sources changes little if the number
of iterations is increased beyond two, or if 
the full exposure correction method (eq.~\ref{eqn:fullcor}) is 
used instead of the fast one (eq.~\ref{eqn:fastcor}).
However, there
is an $\approx$ 5\% increase in source yield by analyzing the image with
the scale sizes spaced by factors of 
$\sqrt{2}$ instead of $2$.  The extra sources 
are relatively weak sources whose probability of detection
is maximized around the scales $\sqrt{2}$ pixels, $2\sqrt{2}$ pixels, etc.
It is the decision of the user as to whether the increase in weak-source
detection efficiency is worth nearly doubling the computation time.

We compare the source detection results for our Test 8 with those
shown in the WGACAT and ROSATSRC catalogs
(White, Giommi, \& Angelini 1994, and Voges et al.~1994, 
respectively),\footnote{
Available from HEASARC: {\tt http://heasarc.gsfc.nasa.gov/}.}
as well as those shown in Micela et al.~and Damiani et al.~(1997b).
(The WGACAT and ROSATSRC catalog teams, and Micela et al., use variants of
the sliding-cell algorithm.)
We choose Test 8 because it 
most closely resembles the analysis of Damiani et al., who
perform a two-iteration analysis of the Pleiades image with
wavelet functions of scale sizes 1, $\sqrt{2}$, ..., 16 pixels,
and with threshold significance $S_o = 1.33\times10^{-5}$ (4.2$\sigma$).
Our result is nearly equal to that of Damiani et al., as 
we detect 148 sources, while they detect 150.
We cannot directly compare our numerical results with those of 
Damiani et al., who only publish figures showing the performance of
their algorithm on a subset of the Pleiades field.
However, we can emulate Damiani et al.~by showing how
our results compare with those of the WGACAT and ROSATSRC catalogs 
(Table \ref{tab:compare}, cf.~Damiani et al.~Table 1).
Our results are virtually identical to those
of Damiani et al., in, e.g.,
how many sources are detected only by our algorithm, etc.
We thus may conclude that our algorithm and that of Damiani 
et al.~generate a largely similar source list.

\begin{figure}
\begin{center}
\includegraphics[scale=0.75]{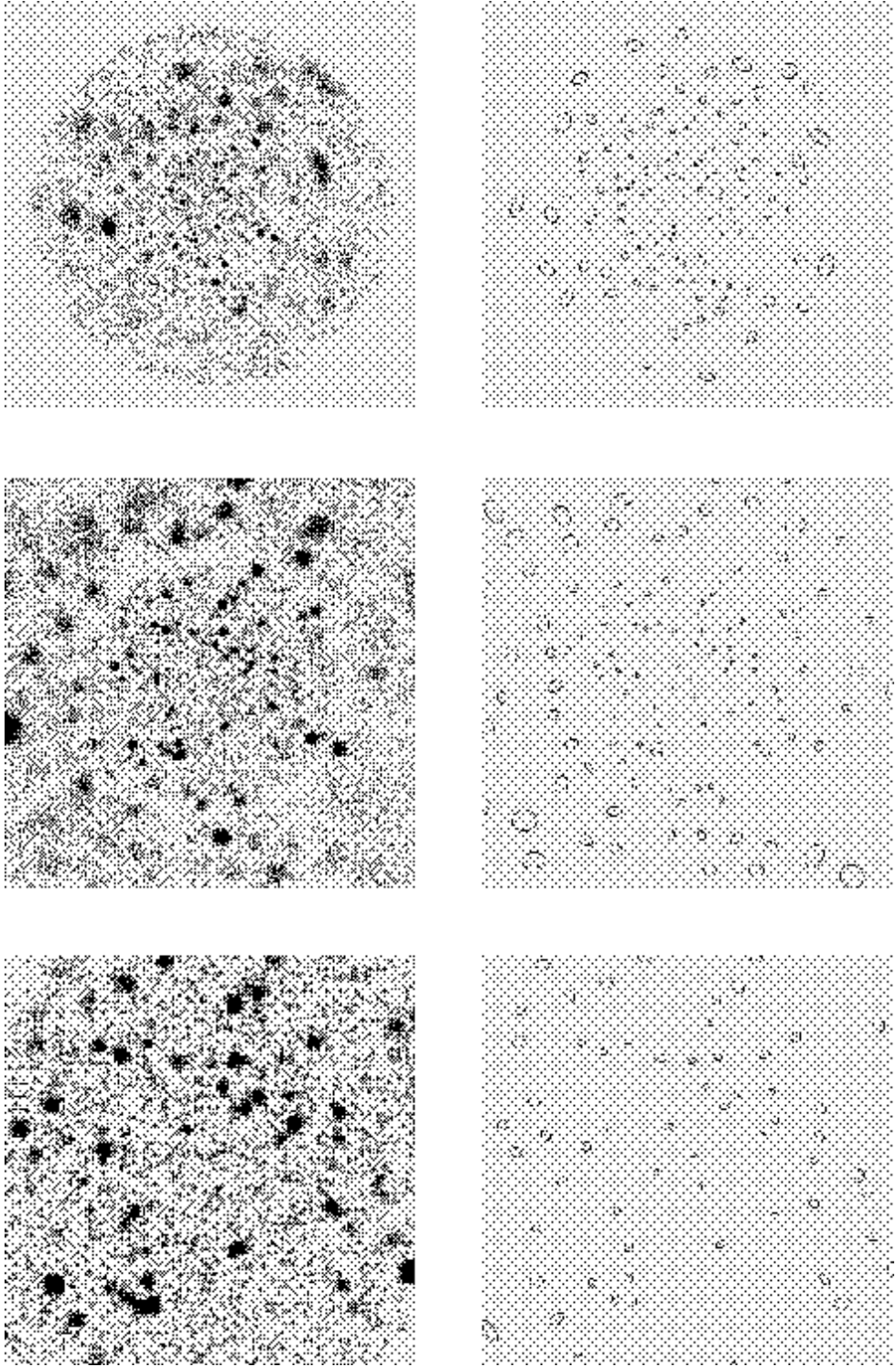}
\end{center}
\caption{
Top Left: the full {\it ROSAT} PSPC image of the Pleiades Cluster.
Top Right: ellipses representing the 148
sources detected by our algorithm.  The ellipse
sizes are set by deriving the 1$\sigma$ principal axes and rotation
angle for each source.
Middle Left and Right:
same as top left and right, with only the central
1$^{\circ}{\times}$1$^{\circ}$ portion of the image shown.
Bottom Left and Right: same as top left and right, with only the central
30$^{\arcmin}{\times}$30$^{\arcmin}$ portion of the image shown.
NOTE: this is a bitmap image.  The original image
may be downloaded from 
{\tt http://hea-www.harvard.edu/{\textasciitilde}pfreeman/f15.eps.gz}.
}
\label{fig:ple}
\end{figure}

A large fraction of the sources that we detect, but that are not included
in the WGACAT or ROSATSRC catalogs, are located either near the inner telescope 
support
ring ($\approx$ 20$^{\arcmin}$ off-axis) or near the precipitous drop-offs
in exposure caused by telescope vignetting near the edge of the FOV
(Figure \ref{fig:ple}; see also Figure \ref{fig:spropest}).  
A visual examination of these sources indicates
that they are not spurious.
We further examine in detail the two sources
near the edge of the FOV that were not included in either the WGACAT or
ROSATSRC catalogs, but are in regions covered by other {\it ROSAT} PSPC
pointings described by Stauffer et al.~(1994).
We find that these
sources lie within $r_{\rm PSF}$ pixels of the Stauffer et al.~sources 159
and 197 (see their Table 2), 
which have reported count rates $\approx$ 0.023 and 0.016 ct 
sec$^{-1}$ respectively.  These rates are consistent with our derived
count rates.
This demonstrates the robustness of our simple exposure correction method.

We also note the intriguing result that the local background map computed by
our algorithm indicates the presence of an X-ray shadow in the core of the
cluster (Figure \ref{fig:ple_bkg}; Kashyap et al.~2001, in preparation).
Because the Pleiades cluster is located beyond the edge of the
local bubble of hot gas (Frisch 1995), this shadow, which is most
pronounced at low energies, is of more distant sources of the diffuse
X-ray background (DXBG), such as the extragalactic component and obscured
stars in the Pleiades cluster itself.  The depth of the shadow places
constraints on the nature of the stellar mass-function at low masses, and
in particular rules out models where the mass-function is extrapolated
at a constant slope from higher masses, thus providing independent,
X-ray observational support for optical observations that report drops
in the mass-function at low masses ($M_B \sim 10$; see, e.g., 
Tinney, Mould, \& Reid 1992, Bahcall et al.~1994).

\subsection{Chandra ACIS-I: The Lockman Hole}

\label{sect:ver2}

Finally, we demonstrate that our algorithm can handle the low background
amplitudes which are characteristic of {\it Chandra} observations, while
continuing to outperform the sliding cell, by applying it to
a simulated 30 ksec {\it Chandra} ACIS-I image of the Lockman Hole
(T.~Gaetz, private communication; Figure \ref{fig:lh_data}).
The ACIS-I is comprised of four 1024$\times$1024 pixel CCDs
configured in a 2$\times$2 square, 
with FOV $\approx$ 17$^{\arcmin}\times$17$^{\arcmin}$
($\approx$ 50 times smaller than that of the {\it ROSAT} PSPC).
Within the ACIS-I field are placed: (1) 12 optically identified
{\it ROSAT} PSPC X-ray sources
catalogued by Schmidt et al.~(1998), including one extended cluster
source;\footnote{
Other detected X-ray sources that lie within the ACIS-I FOV,
but were not optically identified, have not been included.}
(2) $\approx$ 6000 point sources sampled from the
Hasinger et al.~(1998) ${\log}N-{\log}S$ distribution between
10$^{-17}$ and 5$\times$10$^{-15}$ erg cm$^{-2}$ sec$^{-1}$;
and (3) 19500 particle background counts (corresponding to a
particle background rate of 1.5$\times$10$^{-7}$ ct pix$^{-1}$ sec$^{-1}$).

If we assume the input parameters listed for Test 1 in Table
\ref{tab:numdet}, and do not use an exposure map, we detect
171 sources in the full ACIS-I field,
of which four are directly associated with the extended galactic cluster
(Figure \ref{fig:lh_data}).
In Figure \ref{fig:logN}, we show differential
${\log}N-{\log}S$ distributions for both detected, and all,
points sources in the FOV.
On the basis of this figure,
we may conclude that our algorithm will efficiently detect sources
in ACIS-I images with fluxes $\simgreat$ 10$^{-15}$ erg cm$^{-2}$ sec$^{-1}$
(0.5 - 2 keV), and has the ability to detect Poisson sampling fluctuations for
sources with fluxes $\simless$ 10$^{-16}$ erg cm$^{-2}$ sec$^{-1}$.
Our result compares very favorably with the performance of 
{\tt CELLDETECT}, which detects 51 sources with 
$\frac{S}{N} \geq$ 3.  (Further comparison of the source detection
efficiencies of {\tt WAVDETECT} and
{\tt CELLDETECT} is provided by Kim et al.~2001, in preparation.)
We conclude that the
relative superiority of our wavelet detection algorithm,
with respect to the sliding cell, is inversely proportional to the 
background amplitude.

In Figure \ref{fig:offax}, we plot
the offsets of the locations of detected sources from their
actual locations within the FOV.
(These offsets are caused by the asymmetry of the {\it Chandra} PSF,
and their magnitude increases with off-axis angle.)
We note two characteristics of these offsets.  First, the variation
in the offsets does not depend on source strength, signifying that
the source-locating process is insensitive to the strength of the
source.  Second, the observed offsets are significantly smaller
than the expected mean separation between sources ($\approx 50$ pix),
implying that the observed offsets are due to the
asymmetries inherent in the PSF and not due to source misidentifications.
We thus find that the weak sources are as ``well-behaved''
as strong sources (about whose detection and identification there can be
little doubt), and hence infer that even the weakest detected sources
are real.

\section{Comparison with Existing Algorithms}

\label{sect:comp}

The source detection algorithm which we present in this work resembles
algorithms published previously by
Vikhlinin et al.~(1994), Rosati et al.~(1995),
Grebenev et al.~(1995), Damiani et al.~(1997a), and
Lazzati et al.~(1998).
In this section, we highlight 
important differences between our algorithm and
these other algorithms, all of which, having been developed for 
analyzing {\it ROSAT} PSPC data, suffer from inherent limitations that
do not allow them to be directly applied to data from, e.g.,
{\it Chandra}.
We do not discuss how our method of source characterization differs
from those described previously, because these other methods are built
upon the premise that the PSF has Gaussian shape.  Thus
they are simply not directly applicable in situations
where the PSF has a more complex shape.

\subsection{Correlation Image}

Our basic method for computing the correlation images is the same as
that used by Vikhlinin et al., Rosati et al., Grebenev et al.,
and Lazzati et al., with the exception that Rosati et al.~use the
symmetric Morlet wavelet function
\begin{equation}
W_{\rm M}(r)~=~\frac{2}{\sigma^2}\left[e^{-\frac{r^2}{\sigma^2}} - \frac{1}{2}e^{-\frac{r^2}{2\sigma^2}}\right]
\end{equation}
instead of the MH function.  However, these authors do not attempt to correct
for exposure variations, as they focus their attention upon the center of the
{\it ROSAT} PSPC FOV, and they use a fundamentally different method to 
determine source detection thresholds (see below).

The method by which Damiani et al.~correct for exposure and 
detect sources differs substantially from ours.
They first divide the raw data image (which they refer to as the
``photon image") by an exposure map on a pixel-by-pixel basis, to create
the so-called ``count-rate image."  Pixels with relative 
exposure less than a certain amount (e.g.~0.2) are not included,
which introduces a sharp edge in the count-rate image
where telescope vignetting becomes important.
Damiani et al.~correlate the wavelet function with this image,
applying an analytic correction to correlation values near the 
edge (as given in their eq.~12).
Because the data in the count-rate image are
not Poisson-distributed, Damiani et al.~must
convert source detection threshold values, derived from their
background map in the photon image, to values appropriate for the
count-rate image.  They accomplish this by dividing the photon-image
detection thresholds by
an effective exposure time $t_{\rm eff}$,
which is not a source exposure time that can be used to
convert estimated source counts to count rates.
Because Damiani et al.~derive the equation with which they compute 
effective exposure time in
the Gaussian limit (see their Appendix B), their exposure correction method
is effectively limited to the high-counts regime: it cannot be applied as
is to, e.g., typical {\it Chandra} data.

\subsection{Background and Source Detection}

As described in {\S}\ref{sect:alg}, we estimate the local background
counts amplitude in each pixel by assuming that within the negative
annulus of the wavelet, $NW$, there are no source counts
(eq.~\ref{eqn:back2}).  We then use that inferred amplitude to
determine source detection thresholds.  This is a generalization of
the approach used by Vikhlinin et al.,
Rosati et al., Grebenev et al., and Lazzati et al.,
in which the correlation variance
$<W^2{\star}D>$ is used to determine thresholds (e.g.,
$\frac{<W{\star}D>}{\sqrt{<W^2{\star}D>}} \geq$ 3.5).  This approach will
work if the background is locally flat or has a locally constant gradient,
in which case the correlation of wavelet and background is zero.
Such an approach is obviously insufficient for use with the whole FOV
of X-ray detector, where support rib shadows and vignetting will affect
the background, and/or in situations 
where the non-instrumental background varies
markedly (such as in the Pleiades; see Figure \ref{fig:ple_bkg}).

Damiani et al.~use the same basic approach to source detection as we
do, in that they compute a local background amplitude, and use it to
compute source detection thresholds.
Damiani et al.~compute the background map by first smoothing
the raw data with a Gaussian with width $\geq 2\sigma_{\rm PSF,i,j}$.
They then interpolate background estimates within support rib shadows.
At each scale, they compute the median value of the
smoothed background within a square region with side-length
$l~=~4(\sigma^2 + \sigma_{\rm PSF,i,j}^2)$ centered on the given
pixel.  
(We explicitly use $\sigma_{\rm PSF}^2$ instead of
$r_{\rm PSF}^2$ to highlight their assumption of Gaussian
PSF shape.)  
After point sources are detected (using a source cleansing significance
that is 0.2$\sigma$ smaller than their source detection significance), regions 
around them containing 95\% of the counts are masked out, and
a refined background estimate is made by
interpolating over the masked regions.

\section{Summary}

\label{sect:sum}

In this work, we present a generalized
wavelet-based source detection algorithm that, in principle,
can be applied immediately to image data collected by 
any photon counts detector, although it was developed specifically
for the analysis of {\it Chandra X-ray Observatory} image data.
We exclusively use
the Marr, or Mexican Hat, wavelet function in this paper,
but the basic details of our algorithm would be unchanged if we
use other wavelet functions, such as the Haar or Morlet wavelet functions.
Aspects of our algorithm include: (1) the computation of the correlation
of the wavelet function and the data image using either analytic or FFT
methods; (2) the computation of a local, exposure-corrected normalized
(i.e.~flat-fielded) background in each pixel;
(3) its applicability within the low-counts regime, as it does
not require a minimum number of background counts per pixel for the
accurate computation of source detection thresholds;
(4) the correction of those correlation values 
which are affected by large exposure variations within the 
wavelet support (due to, e.g., telescope
support ribs or the edge of the field of view),
using either one of two methods given by eq.~\ref{eqn:fullcor} and
eq.~\ref{eqn:fastcor};
(5) the generation of a source list 
in a manner that does not depend upon the details
of the PSF shape, including the creation of
general, data-dependent source cells for the estimation of source 
properties and filtering of extended source data; and
(6) full error analysis.
In these respects, our algorithm is considerably
more general than the similar wavelet-based methods 
developed by Vikhlinin et al.~(1994), Rosati et al.~(1995),
Grebenev et al.~(1995), Damiani et al.~(1997a,b), Starck \& Pierre (1998),
and Lazzati et al.~(1998).
Nearly all of these methods 
were developed specifically for treating data collected
by the {\it ROSAT} PSPC (Starck \& Pierre 1998 apply their method to data
from the {\it ROSAT} HRI); none except Damiani et al.~attempt
to correct for variations in exposure within the FOV; and all except
for Damiani et al.~assume a flat background across the
region of interest.  The relatively more general method of Damiani et al.~is
limited to analyzing data from detectors which have
PSFs with Gaussian shape, and which have high rates of background
accumulation.
These limitations make the Damiani et al.~approach, as published,
inappropriate for use with, e.g., {\it Chandra} data.

In {\S}\ref{sect:ver}, we demonstrate the robustness of our algorithm by
applying it, {\it without algorithmic changes}, to data collected by an
idealized detector with a spatially invariant Gaussian PSF;
to {\it ROSAT} PSPC data of the crowded field of the Pleiades Cluster;
and to a simulated {\it Chandra} ACIS-I image of the Lockman Hole region.
Collectively,
these test cases indicate that our algorithm: (1) effectively detects and
describes point sources, and can be applied to the study of extended 
sources; (2) does not detect more spurious sources than expected; 
(3) is more sensitive than sliding-cell methods; and (4) has equal 
sensitivity to the Damiani et al.~method for the specific case of
{\it ROSAT} PSPC data.  We find that while we can use the algorithm as 
presented to analyze extended sources, such analysis requires
careful monitoring on the part of the analyst.  Work on generalizing 
this analysis is on-going, and will be reported in a future work.

\acknowledgements

The authors acknowledge the support of 
the CXC Beta Test Site grant and NASA grants
NAG5-3173, NAG5-3189, NAG5-3195, NAG5-3196, NAG5-3831,
NAG5-6755, and NAG5-7226.
We are grateful to T.~Gaetz and A.~Vikhlinin for making available the
simulated ACIS-I image of the Lockman Hole, analyzed in {\S}\ref{sect:ver2}.
We would like to thank R.~C.~Nichol, B.~Holden, J.~Flanagan, T.~Calderwood, 
and A.~Dobryzcki for their assistance with coding issues,
E.~Kolaczyk for useful discussions regarding the computation of covariance,
and F.~Damiani, S.~Sciortino, and A.~Bijaoui for useful discussions.
Finally, we would like to thank the anonymous referees for their
detailed comments and suggestions, which greatly helped the preparation of 
this manuscript.  In particular, our description of the vanishing moments
property of wavelets and its relationship to source detection owes much
to one of these referees.

\appendix

\section{Derivation of Formulae Associated with the Mexican Hat Function}

\label{sect:mh}

The source detection algorithm that we have presented
does not explicitly depend on the details of wavelet function itself,
and thus should be applicable with other, simple, wavelet functions
which have one central positive mode.
Hence we have deferred to this Appendix and the
next the derivation of various formulae
that we use in our algorithm which are valid specifically when using
the Marr, or Mexican Hat (MH), wavelet function.

\subsection{Pixel-by-Pixel Integration of the MH Function}

\label{sect:int}

Before carrying out source detection at given scales ($\sigma_x$,$\sigma_y$),
we must determine the grid of values $W_{i-i',j-j'}$ (see 
\ref{eqn:coranal}).  This is accomplished by
integrating the function $W_m(\frac{x}{\sigma_x},\frac{y}{\sigma_y})$,
defined in eqs.~(\ref{eqn:wave1}-\ref{eqn:wave}), on a pixel-by-pixel basis
within an ellipse with
axes with full lengths 10$\sigma_x$ and 10$\sigma_y$, with values farther
from the origin set to zero:
\begin{eqnarray}
W_{i-i',j-j'}~&=&~\int_{x_1}^{x_2} \int_{y_1}^{y_2} dx dy W_m(\frac{x}{\sigma_x},\frac{y}{\sigma_y}) \\
~&=&~\int_{x_1}^{x_2} dx \int_{y_1}^{y_2} dy \left[ 2 - \frac{x^2}{\sigma_{x}^2} - \frac{y^2}{\sigma_{y}^2} \right] \exp\left(-\frac{x^2}{2\sigma_{x}^2}\right) \exp\left(-\frac{y^2}{2\sigma_{y}^2}\right) \,. \label{eqn:anal}
\end{eqnarray}
$(x,y)$ represent coordinates relative to the center of 
pixel $(i,j)$, and 
and $(x_1,x_2)$ and $(y_1,y_2)$ denote the limits of integration over
pixel $(i',j')$ (Figure \ref{fig:coord}):
\begin{eqnarray}
x_1&=&i'-i-\frac{1}{2}~~~~y_1=j'-j-\frac{1}{2} \nonumber \\
x_2&=&i'-i+\frac{1}{2}~~~~y_2=j'-j+\frac{1}{2} \,. \nonumber
\end{eqnarray}

We expand eq.~(\ref{eqn:anal}) and perform each integral separately.
The first integral in the expansion is computed using
\begin{eqnarray}
\int_{x_1}^{x_2} dx \exp\left(-\frac{x^2}{2\sigma_{x}^2}\right)&=&\sqrt{2}\sigma_x \int_{t_1}^{t_2} dt \exp(-t^2) \nonumber \\
&=&\int_0^{t_2} dt \exp(-t^2) - \int_0^{t_1} dt \exp(-t^2) \nonumber \\
&=&\sqrt{\frac{\pi}{2}} \sigma_x \left[{\rm erf}\left(\frac{x_2}{\sqrt{2}\sigma_x}\right) - {\rm erf}\left(\frac{x_1}{\sqrt{2}\sigma_x}\right)\right] \nonumber \\
&=&\sqrt{\frac{\pi}{2}} \sigma_x x_{\rm erf} \,, \nonumber
\end{eqnarray}
where the substitution $t = \frac{x}{\sqrt{2}\sigma_x}$ is made, and 
${\rm erf}(x)$ is the error function.
Performing the same integration over $y$, we determine that the first
term is $2\frac{\pi}{2}\sigma_x\sigma_yx_{\rm erf}y_{\rm erf} =
{\pi}\sigma_x\sigma_yx_{\rm erf}y_{\rm erf}$.

To compute the second and third integrals, we must determine, e.g.,
\begin{equation}
\int_{x_1}^{x_2} dx \frac{x^2}{\sigma_x^2} \exp\left(-\frac{x^2}{2\sigma_{x}^2}\right) = 2\sqrt{2}\sigma_x \int_{t_1}^{t_2} dt t^2 \exp(-t^2) \nonumber .
\end{equation}
This integral can be solved by parts, by taking the derivative of $t$ and
the integral of $t\exp(-t^2)$.  We present the solution:
\begin{equation}
x_1 \exp\left(-\frac{x_1^2}{2\sigma_{x}^2}\right) - x_2  \exp\left(-\frac{x_2^2}{2\sigma_{x}^2}\right) + \sqrt{\frac{\pi}{2}} \sigma_x x_{\rm erf}~=~x_{\rm diff} + \sqrt{\frac{\pi}{2}} \sigma_x x_{\rm erf} \nonumber .
\end{equation}

The final solution is
\begin{equation}
W_{i-i',j-j'}=\pi\sigma_x\sigma_y{x_{\rm erf}}{y_{\rm erf}} - \sqrt{\frac{\pi}{2}} \sigma_y y_{\rm erf}\left(x_{\rm diff} + \sqrt{\frac{\pi}{2}} \sigma_x x_{\rm erf}\right) - \sqrt{\frac{\pi}{2}} \sigma_x x_{\rm erf}\left(y_{\rm diff} + \sqrt{\frac{\pi}{2}} \sigma_y y_{\rm erf}\right) .
\label{eqn:anlw}
\end{equation}

\subsection{Fourier Transform of the MH Function}

\label{sect:ftwave}

Analytic computation of correlation values (eq.~\ref{eqn:coranal}) may be
too computationally intensive if the image and/or wavelet scale sizes are
too large.  So in {\tt WAVDETECT}, for instance,
Fast Fourier Transforms\footnote{
The CIAO {\tt WAVDETECT} code (written in C) makes use of the FFTW
package (Frigo \& Johnson 1998), while our own Fortran code uses the
publicly available MFFT algorithm (Nobile \& Roberto 1986).} 
(FFTs) are used if scale sizes are $\geq$ 2 pixels:
\begin{equation}
C~\approx~{\rm FFT}^{-1} \left[N \times {\rm FFT}(W) \times {\rm FFT}^{\ast}(D) \right] .
\end{equation}
To mitigate the effect of the ``wrap-around,"
any image that is to be transformed is padded with zeros;
the minimum width of the padding is 10${\times}{\rm max}(\sigma_x,\sigma_y)$.

In our own Fortran code, we use the analytic Fourier Transform 
of the MH function, which we denote $FT(W)$:
\begin{eqnarray}
FT(W)~&=&~\int_{-\infty}^{+\infty} \int_{-\infty}^{+\infty} dx dy W(x,y) e^{2{\pi}i(k_{x}x+k_{y}y)} \nonumber \\
&=&~\int_{-\infty}^{+\infty} dy e^{2{\pi}ik_{y}y} \int_{-\infty}^{+\infty} dx W(x,y) \cos(2{\pi}k_{x}x) \nonumber \\
&&~+~i \int_{-\infty}^{+\infty} dy e^{2{\pi}ik_{y}y} \int_{-\infty}^{+\infty} dx W(x,y) \sin(2{\pi}k_{x}x) \label{eqn:evodd} \\
&=&~\int_{-\infty}^{+\infty} dy e^{2{\pi}ik_{y}y} \int_{-\infty}^{+\infty} dx W(x,y) \cos(2{\pi}k_{x}x) \nonumber \\
&=&~\left[\int_{-\infty}^{+\infty} dy \cos(2{\pi}k_{y}y) + i\int_{-\infty}^{+\infty} dy \sin(2{\pi}k_{y}y)\right] \int_{-\infty}^{+\infty} dx W(x,y) \cos(2{\pi}k_{x}x) \label{eqn:evodd2} \nonumber \\
&=&~\int_{-\infty}^{+\infty} dy \cos(2{\pi}k_{y}y) \int_{-\infty}^{+\infty} dx W(x,y) \cos(2{\pi}k_{x}x) \nonumber \\
&=&~\int_{-\infty}^{+\infty} dy \cos(2{\pi}k_{y}y) \times C_{\rm W} \label{eqn:cw} \,.
\end{eqnarray}
The wave-number $k$ equals $\frac{2(i_k-1)k_{\rm N}}{l}$, where
$i_k$ is the pixel number in Fourier space, $k_{\rm N} = \frac{1}{2}$ is
the Nyquist wave-number, and $l$ is half the length of the relevant axis in
the padded image.
The fourth integral in eq.~(\ref{eqn:evodd}) and the second integral in 
eq.~(\ref{eqn:evodd2}) are zero, as the integrands are
products of even and odd functions.
\begin{eqnarray}
C_{\rm W}~&=&~\exp\left(-\frac{y^2}{2\sigma_y^2}\right) \int_{-\infty}^{+\infty} dx \left[2 - \frac{x^2}{\sigma_x^2} -\frac{y^2}{\sigma_y^2}\right] \exp\left(-\frac{x^2}{2\sigma_x^2}\right)\cos(2{\pi}k_xx) \nonumber \\
&=&~\exp\left(-\frac{y^2}{2\sigma_y^2}\right) \left[ \left(2 - \frac{y^2}{\sigma_y^2} \right) \Psi_{\rm 0,c}(2{\pi}k_x,\frac{1}{\sqrt{2}\sigma_x},0) - \frac{1}{\sigma_x^2} \Psi_{\rm 2,c}(2{\pi}k_x,\frac{1}{\sqrt{2}\sigma_x}) \right] \label{eqn:c2} \,.
\end{eqnarray}
$\Psi_{\rm 0,c}(p,q,{\lambda})$ and $\Psi_{\rm 2,c}(a,p)$ represent two
integral solutions which one may find, e.g., in Gradshteyn \& Ryzhik~(1980;
formulae 3.896-2 and 3.952-4 respectively):
\begin{equation}
\Psi_{\rm 0,c}(p,q,{\lambda})~=~\frac{1}{q}\sqrt{\pi}\exp\left(-\frac{p^2}{4q^2}\right)\cos(p{\lambda}) \nonumber ,
\end{equation}
and
\begin{equation}
\Psi_{\rm 2,c}(a,p)~=~\sqrt{\pi}\frac{2p^2-a^2}{4p^5}\exp\left(-\frac{a^2}{4p^2}\right) \nonumber .
\end{equation}
Substituting these quantities into eq.~(\ref{eqn:c2}), one finds that
\begin{equation}
C_{\rm W}~=~\exp\left(-\frac{y^2}{2\sigma_{y}^2}-2{\pi}^2k_{x}^2\sigma_{x}^2\right) \sigma_{x} \sqrt{2{\pi}} \left[ 4{\pi}^2k_{x}^2\sigma_{x}^2 + 1 - \frac{y^2}{\sigma_{y}^2} \right]
\label{eqn:cw2}
\end{equation}
Substituting eq.~(\ref{eqn:cw2}) into eq.~(\ref{eqn:cw}) and 
solving, we find:
\begin{eqnarray}
&&FT(W) \nonumber \\
&=&~\exp(-2{\pi}^2k_{x}^2\sigma_{x}^2) \sigma_{x} \sqrt{2{\pi}} \int_{-\infty}^{+\infty} dy \cos(2{\pi}k_{y}y) \exp\left(-\frac{y^2}{2\sigma_{y}^2}\right) \left[ 4{\pi}^2k_{x}^2\sigma_{x}^2 + 1 - \frac{y^2}{\sigma_{y}^2} \right] \nonumber \\
&=&~\exp(-2{\pi}^2k_{x}^2\sigma_{x}^2) \sigma_{x} \sqrt{2{\pi}} \left[ (4{\pi}^2k_{x}^2\sigma_{x}^2 + 1)\Psi_{\rm 0,c}(2{\pi}k_{y},\frac{1}{\sqrt{2}\sigma_{y}},0) - \frac{1}{\sigma_{y}^2}\Psi_{\rm 2,c}(2{\pi}k_{y},\frac{1}{\sqrt{2}\sigma_{y}})\right] \nonumber \\
&=&~(2{\pi})^3\sigma_{x}\sigma_{y}(k_{x}^2\sigma_{x}^2 + k_{y}^2\sigma_{y}^2)\exp[-2{\pi}^2(k_{x}^2\sigma_{x}^2+k_{y}^2\sigma_{y}^2)] \,. \label{eqn:ftw}
\end{eqnarray}

By using $FT(W)$ instead of $FFT(W)$, we reduce computation time,
but numerical estimates are less accurate
(with respect to estimates derived analytically).
We quantify the discrepancy between any two of our three methods
($FT(W)$, $FFT(W)$, or analytic) using
\begin{equation}
\delta~=~\frac{\vert {\Delta}C \vert}{T_{\rm an}} .
\end{equation}
We use the analytically derived detection threshold $T_{\rm an}$
in the denominator because the expectation value
of $C$ is zero.  It also provides an intuitive way to describe
the discrepancy at the detection threshold.
We find $\delta_{\rm FFT,an} \sim 10^{-4}$, while
$\delta_{\rm FT,an} \simgreat 10^{-2}$, increasing with
$\frac{C}{T_{\rm an}}$.  (See Figure \ref{fig:ftan}.)

The discrepancy results from the fact that we analyze
binned images.
Because the data are binned, we should actually compute $FT(W)$ 
using the equation
\begin{equation}
FT(W)~=~\int_{-\infty}^{+\infty} \int_{-\infty}^{+\infty} dx dy W_{i-i',j-j'} e^{2{\pi}i(k_{x}x+k_{y}y)} ,
\label{eqn:aft}
\end{equation}
If $W(x,y)$ exhibits significant
curvature in a bin, then $W_{i-i',j-j'} \neq W(x,y)$ within that
bin; the effect of this 
discrepancy in the final correlation value
is proportional to the number of counts in the bin.
Thus ${\delta}_{\rm FT,An}$ will increase with source strength, as
shown in the right panel of Figure \ref{fig:ftan}.

If we wish to
compute the error of the correlation value in a pixel, we calculate
$<W^2{\star}D>_{i,j}$, saving computation time by
using the analytic Fourier Transform $FT(W^2)$.
The derivation of this function is similar to
the derivation of $FT(W)$.  A new integral which appears is:
\begin{equation}
\int_{-\infty}^{+\infty} dx \frac{x^4}{\sigma_{x}^4} \exp\left(-\frac{x^2}{\sigma_{x}^2}\right) \cos(2{\pi}k_{x}x)
\end{equation}
One may solve this integral by parts, differentiating the term
$\frac{x^3}{\sigma_{x}^2}\cos(2{\pi}k_{x}x)$ and integrating the term
$\frac{x}{\sigma_{x}^2}\exp\left(-\frac{x^2}{\sigma_{x}^2}\right)$.  The
integral
\begin{equation}
\int_{-\infty}^{+\infty} dx x^3 \sin(2{\pi}k_{x}x) \exp\left(-\frac{x^2}{\sigma_{x}^2}\right)
\end{equation}
has solution (see, e.g., Gradshteyn \& Ryzhik 1980, formula 3.952-5):
\begin{equation}
\int_{-\infty}^{+\infty} dx x^3 \sin(2{\pi}k_{x}x) \exp\left(-\frac{x^2}{\sigma_{x}^2}\right)~=~\sqrt{\pi}\sigma_{x}^7\frac{3{\pi}\frac{k_{x}}{\sigma_{x}^2}-2{\pi}^3k_{x}^3}{2}\exp(-{\pi}^2\sigma_{x}^2k_{x}^2) .
\end{equation}

The final solution is
\begin{equation}
FT(W^2)~=~\left[2{\pi}\sigma_{x}\sigma_{y} + {\pi}^5\sigma_{x}\sigma_{y}(k_{x}^2\sigma_{x}^2+k_{y}^2\sigma_{y}^2)^2\right]\exp[-{\pi}^2(k_{x}^2\sigma_{x}^2+k_{y}^2\sigma_{y}^2)] .
\end{equation}

Use of $FT(W^2)$ instead of $FFT(W^2)$ leads to a less accurate accounting
of the errors as derived using purely analytic methods, for reasons
given above.  We find the magnitude of the discrepancy in error
estimates, as a fraction of the detection threshold, to be similar to that
seen above ($\sim$ 10$^{-2}$); if we use the analytic error in the
denominator instead of the detection threshold, we find that the discrepancy
is constant as a function of $\frac{C}{T}$, at $\sim$ 10$^{-2}$.

\subsection{Integration of the MH Function Negative Annulus}

\label{sect:negann}

If exposure variations and the FOV edge may be ignored in the computation of
the background, then we may replace $<NW{\star}E>_{i,j}$ in
eq.~(\ref{eqn:back2}) with a background 
normalization factor, $N_B$, which we derive here.

The average value of the MH function is zero.  
Hence $N_B$ may be derived by integrating the MH function
over either $PW$ or $NW$.  We choose the former:
\begin{equation}
N_B~=~\int_{-\sqrt{2}\sigma_{x}}^{\sqrt{2}\sigma_{x}} dx \int_{-[2-\frac{x^2}{\sigma_{x}^2}]^{\frac{1}{2}}\sigma_y}^{[2-\frac{x^2}{\sigma_{x}^2}]^{\frac{1}{2}}\sigma_y} dy \left[ 2 - \frac{x^2}{\sigma_{x}^2} - \frac{y^2}{\sigma_{y}^2} \right] \exp\left(-\frac{x^2}{2\sigma_{x}^2}\right) \exp\left(-\frac{y^2}{2\sigma_{y}^2} \right) .
\end{equation}
The limits of integration reflect that the core extends over an
ellipse with axes $\sqrt{2}\sigma_x$ and $\sqrt{2}\sigma_y$.
We reparametrize the integral using
polar coordinates $(r,\theta)$,  
after first mapping the boundary ellipse to a boundary circle using the
transformation $y' = \frac{\sigma_{x}}{\sigma_{y}}y$:
\begin{eqnarray}
N_B~&=&~\frac{\sigma_{y}}{\sigma_{x}} \int_{-\sqrt{2}\sigma_{x}}^{\sqrt{2}\sigma_{x}} dx \int_{-\sigma_{x}\sqrt{2-\frac{x^2}{\sigma_{x}^2}}}^{\sigma_{x}\sqrt{2-\frac{x^2}{\sigma_{x}^2}}} dy' \left[ 2 - \frac{x^2+y'^2}{\sigma_{x}^2} \right] \exp\left(-\frac{x^2+y'^2}{2\sigma_{x}^2}\right) \\
&=&~\frac{\sigma_{y}}{\sigma_{x}}\int_0^{2{\pi}} d\theta \int_0^{\sqrt{2}\sigma_{x}} dr r \left[2 - \frac{r^2}{\sigma_{x}^2}\right] \exp\left(-\frac{r^2}{2\sigma_{x}^2}\right) \\
&=&~2{\pi}\frac{\sigma_{y}}{\sigma_{x}} \left[\int_0^{\sqrt{2}\sigma_{x}} dr 2 r \exp\left(-\frac{r^2}{2\sigma_{x}^2}\right) - \int_0^{\sqrt{2}\sigma_{x}} dr \frac{r^3}{\sigma_{x}^2} \exp\left(-\frac{r^2}{2\sigma_{x}^2}\right)\right] \,. \label{eqn:1back}
\end{eqnarray}
The determinant of the Jacobian of the transformation from 
$(x,y') \rightarrow (r,\theta)$ is $r$.

We evaluate the second integral using integration by parts:
\begin{equation}
\int_0^{\sqrt{2}\sigma_{x}} dr \frac{r^3}{\sigma_{x}^2} \exp\left(-\frac{r^2}{2\sigma_{x}^2}\right)~=~-\left[r^2\exp\left(-\frac{r^2}{2\sigma_{x}}\right)\mid_0^{\sqrt{2}\sigma_{x}}\right] + \int_0^{\sqrt{2}\sigma_{x}} dr 2 r \exp\left(-\frac{r^2}{2\sigma_{x}^2}\right) .
\label{eqn:2back}
\end{equation}
The second integral in eq.~(\ref{eqn:2back}) cancels with the first integral
in eq.~(\ref{eqn:1back}), leaving:
\begin{eqnarray}
N_B~&=&~2{\pi}\frac{\sigma_{y}}{\sigma_{x}} \left[r^2\exp\left(-\frac{r^2}{2\sigma_{x}}\right)\mid_0^{\sqrt{2}\sigma_{x}}\right] \nonumber \\
&=&~2{\pi}\frac{\sigma_{y}}{\sigma_{x}} \frac{2\sigma_{x}^2}{\exp(1)} \nonumber \\
&=&~\frac{4{\pi}\sigma_{x}\sigma_{y}}{\exp(1)} \,.
\end{eqnarray}

\section{Computation of Detection Thresholds}

\label{sect:sims}

We associate an image pixel $(i,j)$
with a source if the significance $S_{i,j}$
of its correlation value $C_{i,j}$ is greater
than a user-defined detection threshold significance $S_o$:
\begin{equation}
S_{i,j}~=~\int_{C_{i,j}}^{\infty} dC p(C \vert B_{i,j}) > S_o , \nonumber
\end{equation}
where $B_{i,j}$ is either $<NW{\star}D>_{i,j}$, the convolution of the wavelet
negative annulus $NW$ with (possibly cleansed) image data $D$, or the
amplitude in an input background map.
$p(C \vert B_{i,j})$ is the probability sampling distribution (PSD) for
$C$ given $B_{i,j}$.
(We are assuming here that the field of view is evenly exposed, which is
true for the simulations we describe below.
In a more general expression, we would replace
$<NW{\star}D>_{i,j}$ with 
$E_{i,j}\frac{<NW{\star}D>_{i,j}}{{<NW{\star}E>_{i,j}}}$.)
By defining the significance in this manner, we are making the assumption
that the expected background counts amplitude is constant throughout the
wavelet support, and that
if we are computing the background, there
are no source counts in the negative annulus.
Since the PSD is dimensionless, it is independent
of scale size (e.g.~the PSD for $B_{i,j}$
would be the same if we were to double $\sigma_x$ and $\sigma_y$
and reduce the count rate by a factor of four).

We replace $B_{i,j}$ with the equivalent quantity
$q_{i,j}$, the expected number of background counts within the spatial
region spanned by the positive kernel of the wavelet, $PW$.
($q_{i,j}$ is related to
$B_{i,j}$ by multiplicative factors.)
The PSD does not have analytic form when the total number of expected counts
within the positive kernel of the wavelet, $q_{i,j}$, is small ($\simless$ 1);
this is demonstrated by Damiani et al.~(1997a).  Like Damiani et al., we use
simulations to estimate significances and detection thresholds; we carry
these simulations out both within the regime they examine and also for
lower expected counts values.
Because this is a computationally intensive problem,
we cannot accurately estimate significances for each pixel
if $S_{i,j} \simless 10^{-7}$.
Instead, we compare the value of $C_{i,j}$ in each pixel with the
detection threshold $C_{i,j,o}(S_o,B_{i,j})$, defined by
\begin{eqnarray}
S_o~&=&~\int_{C_{i,j,o}}^{\infty} dC p(C \vert B_{i,j}) \nonumber \\
&=&~\int_{C_{i,j,o}}^{\infty} dC p(C \vert q_{i,j}) \,. \nonumber 
\end{eqnarray}
Here, we recast the equation using the variable $q_{i,j}$, the number of
expected counts in $PW$, as its use clarifies our description below 
of how we estimate $C_{i,j,o}$.

To determine $C_{i,j,o}(S_o,q_{i,j})$, we simulated over 50,000 
1024$\times$1024 flat-field images.\footnote{Images that had at least
one count; empty images were ignored.} For each image, we:
\begin{itemize}
\item randomly selected values ${\log}q_o$ from the range
-10 $\leq {\log}q_o \leq$ 3.25 (we describe why we chose this upper
limit below);
\item determined the expected background amplitude
$B_o = \frac{q_o}{2{\pi}\sigma^2}$ in each image pixel 
($\sigma = 4$ pixels);
\item sampled data in each pixel from the appropriate
Poisson distribution given rate $B_o$; 
\item computed $C_{i,j}$ and $q_{i,j}$ for each image pixel;
\item and recorded $C_{i,j}$ in bins of size ${\Delta}({\log}q_{i,j})$
= 0.2, for -6.9 $\leq {\log}q_{i,j} \leq$ 3.1, 
with one bin being used for all values of ${\log}q_{i,j} <$ -6.9.\footnote{
This is of the order of the machine precision.}
From these distributions $p(C \vert q_{i,j})$, we can determine $C_{i,j,o}$.
\end{itemize}
Because there is an inverse correlation between observed values of
$C_{i,j}$ and $q_{i,j}$, it is important to record
$p(C \vert q_{i,j})$ and {\it not} $p(C \vert q_o)$.  Use of the latter
distribution leads to underestimated detection thresholds, and
thus to larger numbers of false source detections than one would 
expect, given $S_o$.

We determined 25 values of $C_o(S_o,q)$ in each ${\log}q$ bin,
for values of $S_o \simgreat$ 10$^{-7}$, using
the central 68\% (17 values) to estimate ``one-$\sigma$" errors on $C_o$.
We then fit these data with simple functions, minimizing the $\chi^2$ statistic.
These functions we use describe the observed detection thresholds well, except
in the regime ${\log}q_{i,j} \simless 0$ and ${\log}S_o \simgreat -4$,
where we use a look-up table instead.
These fits allow us in principle
to compute detection threshold for significances below
our computational lower limit $S_o \sim$ 10$^{-7}$ (such as for
$S_o \sim$ 10$^{-9}$, the significance corresponding to one false source
pixel in an {\it Chandra} HRC image).

In the regime ${\log}q_{i,j} \simless 0$ and ${\log}S_o \simless -4$,
we compute ${\log}C_o(q_{i,j})$ using the function
\begin{equation}
{\log}C_o(q_{i,j})~=~A_{\rm lo}({\log}q_{i,j})^2 + B_{\rm lo}({\log}q_{i,j}) + C_{\rm lo} ,
\end{equation}
where
\begin{eqnarray}
A_{\rm lo}~&=&~0.00462 \nonumber \\
B_{\rm lo}~&=&~0.0661 \nonumber \\
C_{\rm lo}~&=&~-0.0154({\log}S_o)^2 - 0.252({\log}S_o) - 0.031 \,. \nonumber
\end{eqnarray}

In the regime ${\log}q_{i,j} \simgreat 0$, we compute $C_o(q_{i,j})$ 
using the function
\begin{equation}
C_o(q_{i,j})~=~A_{\rm hi}\sqrt{q_{i,j}} + B_{\rm hi} ,
\label{eqn:hi}
\end{equation}
where
\begin{eqnarray}
A_{\rm hi}~&=&~\cases{-0.509({\log}S_o) + 1.897 - .00172~({\log}S_o+7)^{3.606} & ${\log}S_o \geq$ -7 \cr
-0.509({\log}S_o) + 1.897 & ${\log}S_o <$ -7} \nonumber \\
B_{\rm hi}~&=&~-1.115({\log}S_o) - 1.038 \,. \nonumber
\end{eqnarray}
This function is also used by Damiani et al.~to fit detection
thresholds in this counts regime, though their derived coefficients differ
from ours.

For values ${\log}q_{i,j} \sim 0$, we find that we must use
the formula
\begin{equation}
{\log}C_o(q_{i,j})~=~A_{\rm mid}({\log}q_{i,j}) + B_{\rm mid}
\end{equation}
to compute the detection threshold, with
\begin{equation}
A_{\rm mid}~=~0.00182({\log}S_o)^3 + 0.0279({\log}S_o)^2 + 0.158({\log}S_o) + 0.607
\end{equation}
and
\begin{equation}
B_{\rm mid}~=~\cases{
-0.064({\log}S_o) + 0.612 - 0.000115({\log}S_o+7)^{4.75} & -2 $< {\log}S_o \leq$ -1\cr
-0.064({\log}S_o) + 0.612 - 0.00085({\log}S_o+7)^{3.5} & -7 $\leq {\log}S_o \leq$ -2\cr
-0.064({\log}S_o) + 0.612 & ${\log}S_o <$ -7\cr
}
\end{equation}

For our simulations, we chose ${\log}q_o =$ 3.25 
as a upper limit because of the contention of
Damiani et al.~that if ${\log}q_{i,j} \simgreat$ 3, the PSD
probability sampling distribution is analytically representable as
a Gaussian with width $\sigma = \sqrt{q_{i,j}}$.  
(Note that the value of
$q_{i,j}$ that we use in this work is larger than that used by Damiani
et al.~by a factor of $2{\pi}$.)
If this is the case, the significance is given by:
\begin{eqnarray}
S_{i,j}~&=&~\frac{1}{\sqrt{2{\pi}q_{i,j}}}\int_{C_{i,j}}^{\infty} dC \exp\left(-\frac{C_{i,j}^2}{2q_{i,j}}\right) \nonumber \\
&=&~1 - \frac{1}{2} - \frac{1}{\sqrt{2{\pi}q_{i,j}}}\int_0^{C_{i,j}} dC \exp\left(-\frac{C_{i,j}^2}{2q_{i,j}}\right) \nonumber \\
&=&~\frac{1}{2} \left[1 - {\rm erf}\left(\frac{C_{i,j}}{\sqrt{2q_{i,j}}}\right)\right] \,.
\label{eqn:anl}
\end{eqnarray}
We find, however, that if we use this formula in the regime
${\log}q_{i,j} \simgreat$ 3, the derived values of $C_{i,j,o}$ are smaller
than those predicted by eq.~(\ref{eqn:hi}) above.
Thus, because it is more conservative, we use eq.~(\ref{eqn:hi})
to compute detection thresholds for all values ${\log}q_{i,j} \simgreat$ 0.

\section{Covariance Estimate: Two-Iteration Background}

\label{sect:covar}

If the data are cleansed (see {\S}\ref{sect:back}), then an exact
calculation of $V[B_{i,j}]$ will include non-zero covariance terms.
In this section, we derive these terms assuming that the data are
cleansed only once, i.e.~we compute the variance for $B_{2,i,j}$,
a background estimate made by convolving the wavelet negative annulus
with data $D_2$ that are a mixture of raw data $D$ and first-iteration
background estimates $B_1$:
\begin{equation}
B_{2,i,j}~=~E_{i,j} \frac{<NW{\star}D_2>_{i,j}}{<NW{\star}E>_{i,j}}~=~N_{i,j} \sum_{i'}\sum_{j'} NW_{i-i',j-j'}D_{i',j'}' \,,
\end{equation}
where $N_{i,j}$ = $E_{i,j}{\slash}<NW{\star}E>_{i,j}$.

Using eq.~(\ref{eqn:covar}),
\begin{eqnarray}
V[B_{2,i,j}]~&=&~\sum_{i'} \sum_{j'} (N_{i,j}NW_{i-i',j-j'})^2 V[D_{i',j'}'] \nonumber \\
&&+ 2 \sum_{i'} \sum_{i'' > i'} \sum_{j'} \sum_{j'' > j'} (N_{i,j}NW_{i-i',j-j'}) (N_{i,j}NW_{i-i'',j-j''}) {\rm cov}[D_{i',j'}',D_{i'',j''}'] \,. \nonumber
\end{eqnarray}
To calculate the variance, we must estimate both $V[D_{i',j'}']$ and
${\rm cov}[D_{i',j'}',D_{i'',j''}']$.  The estimate of the former
is given in eq.~(\ref{eqn:twoiter}), derived
assuming that the raw datum $D_{i',j'}$ is sampled from
a Poisson distribution with variance $D_{i',j'}$, and that each pixel's
raw datum is independently sampled.

Estimation of ${\rm cov}[D_{i',j'}',D_{i'',j''}']$ is more complicated.
For the two-iteration background case, there are three possibilities:
$D_{i',j'}' = D_{i',j'}$ and $D_{i'',j''}' = D_{i'',j''}$;
$D_{i',j'}' = D_{i',j'}$ and $D_{i'',j''}' = B_{1,i'',j''}$ (or
vice-versa); or
$D_{i',j'}' = B_{1,i',j'}$ and $D_{i'',j''}' = B_{1,i'',j''}$.
Assuming that the raw data are independently sampled, in the first
case the covariance is zero.  In the second case, we can estimate
the covariance using the approximation (Eadie et al.~p.~27):
\begin{eqnarray}
{\rm cov}[D_{i',j'},B_{1,i'',j''}]~&=&~\sum_k \sum_l \sum_{k'} \sum_{l'} \left(\frac{{\partial}D_{i',j'}}{{\partial}\mu_{k,l}}\right) \left(\frac{{\partial}B_{1,i'',j''}}{{\partial}\mu_{k',l'}}\right) cov[D_{k,l},D_{k',l'}] \nonumber \\
&=&~\sum_k \sum_l \left(\frac{{\partial}D_{i',j'}}{{\partial}\mu_{k,l}}\right) \left(\frac{{\partial}B_{1,i'',j''}}{{\partial}\mu_{k,l}}\right) V[D_{k,l}] \nonumber \\
&=&~\sum_k \sum_l \left(\frac{{\partial}D_{i',j'}}{{\partial}D_{k,l}}\right) \left(\frac{{\partial}B_{1,i'',j''}}{{\partial}D_{k,l}}\right) V[D_{k,l}] \nonumber \\
&=&~\left(\frac{{\partial}B_{1,i'',j''}}{{\partial}D_{i',j'}}\right) V[D_{i',j'}] \nonumber \\
&=&~V[D_{i',j'}] \frac{\partial}{{\partial}D_{i',j'}} N_{i'',j''} \sum_{i'''} \sum_{j'''} NW_{i''-i''',j''-j'''} D_{i''',j'''} \nonumber \\
&=&~V[D_{i',j'}] N_{i'',j''} NW_{i''-i',j''-j'} \nonumber \\
&=&~D_{i',j'} N_{i'',j''} NW_{i''-i',j''-j'} \nonumber \,.
\end{eqnarray}
In the above equations, $\mu$ represents the expectation value of
the sampling distribution for $D$; for a Poisson distribution with
variance equal to the datum, $\mu$ will be equal to the datum.
In the third case, we find (while skipping some steps):
\begin{eqnarray}
{\rm cov}[B_{1,i',j'},B_{1,i'',j''}]~&=&~\sum_k \sum_l \left(\frac{{\partial}B_{1,i',j'}}{{\partial}D_{k,l}}\right) \left(\frac{{\partial}B_{1,i'',j''}}{{\partial}D_{k,l}}\right) V[D_{k,l}] \nonumber \\
&=&~\sum_k \sum_l N_{i',j'} NW_{i'-k,j'-l} N_{i'',j''} NW_{i''-k,j''-l} V[D_{k,l}] \nonumber \\
&=&~\sum_k \sum_l N_{i',j'} NW_{i'-k,j'-l} N_{i'',j''} NW_{i''-k,j''-l} D_{k,l} \nonumber \,.
\end{eqnarray}

To demonstrate the effect of including covariance terms, we compute
$V[B_2]$ for a 50$\times$50 subfield of the {\it ROSAT} PSPC Pleiades image,
with $\sigma_x$ = $\sigma_y$ = 4 pixels.
In Figure \ref{fig:covar}, we show the ratio 
$V[B_2]_{\rm covar}{\slash}V[B_2]_{\rm no covar}$.
We infer the following from this computation:
(1) it is quantitatively unimportant at the location of sources and within
voids, where the change in variance is $\simless$ 1\%; (2) in the vicinity of
a strong sources, the change in variance is $\sim$ 10\% (with maximum
$\approx$ 30\% for this particular field, in the vicinity of an 800-count
source), and is affected by source crowding; and 
(3) including covariance terms increases the
CPU time needed to compute $V[B_2]$ by a factor
$\sim {\cal O}(d_xd_y\sigma_x^2\sigma_y^2)$, where $d_x$ and $d_y$ are the
x- and y-axis lengths, respectively, in pixels.

\vfill\eject

\newpage

\clearpage

\clearpage

\clearpage

\clearpage

\clearpage

\clearpage

\clearpage

\clearpage

\clearpage

\clearpage

\clearpage

\clearpage

\clearpage

\clearpage

\clearpage
\begin{figure}
\begin{center}
\includegraphics[angle=-90,scale=0.60]{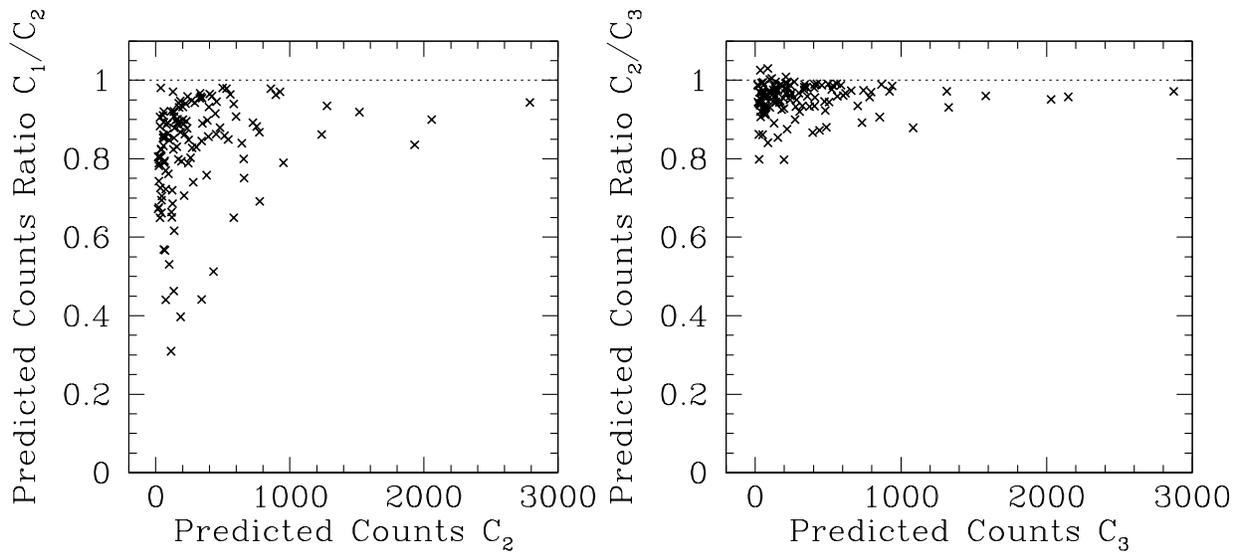}
\end{center}
\caption{
Illustration of how the number of iterations used to estimate
the local background amplitude ({\S}\ref{sect:back}) affects
the estimation of Pleiades Cluster source properties ({\S}\ref{sect:ver3}).
Left: the ratio of source predicted counts $C_1{\slash}C_2$ given backgrounds
estimated using one iteration, and two iterations, as a function of $C_2$.
Right: same as left, but instead computed for two and three iterations.
}
\label{fig:spropest}
\end{figure}

\clearpage
\begin{figure}
\begin{center}
\includegraphics[scale=0.75]{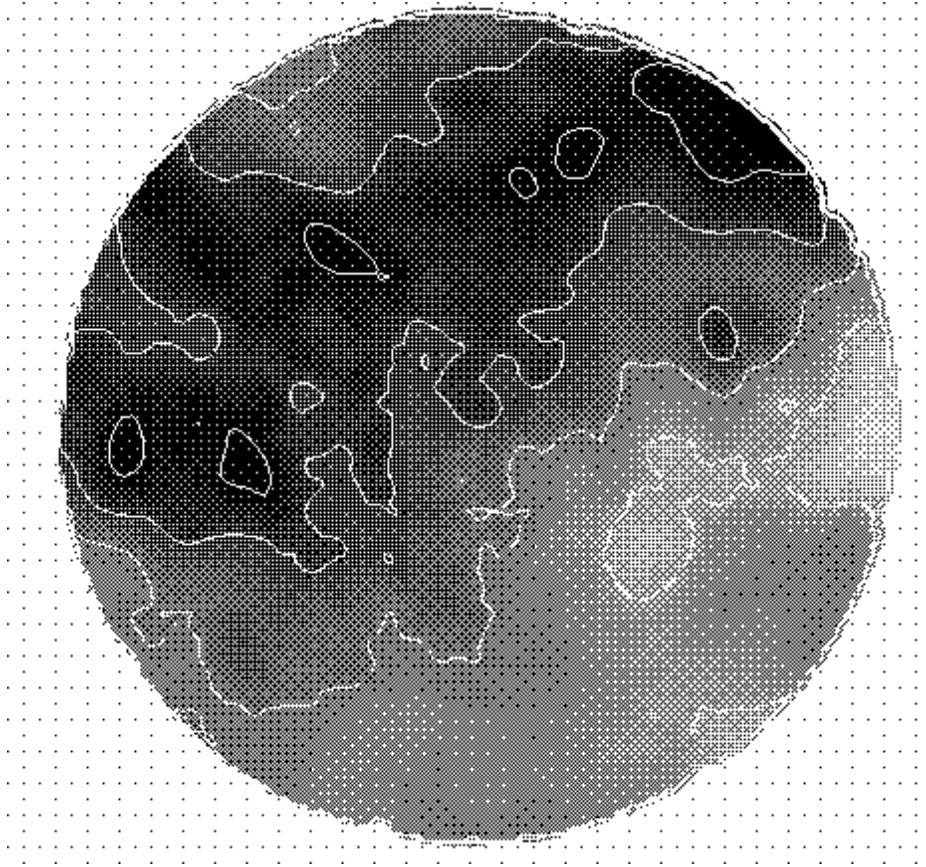}
\end{center}
\caption{
Corrected background map generated during the analysis of the 
{\it ROSAT} PSPC image of the Pleiades Cluster ({\S}\ref{sect:ver3})
via the method of {\S}\ref{sect:nbkg}.  The contour levels are 
0, 0.9, 1.05, 1.2, 1.35, and 1.5 counts, with darker areas having
more counts.  Since the estimated error in each
pixel is $\sim$ 0.01 counts, the perceived structure is real and is
indicative of X-ray shadowing (Kashyap et al.~2001, in preparation).
NOTE: this is a bitmap image.  The original image
may be downloaded from 
{\tt http://hea-www.harvard.edu/{\textasciitilde}pfreeman/f17.eps.gz}.
}
\label{fig:ple_bkg}
\end{figure}

\clearpage
\begin{figure}
\begin{center}
\includegraphics[scale=0.75]{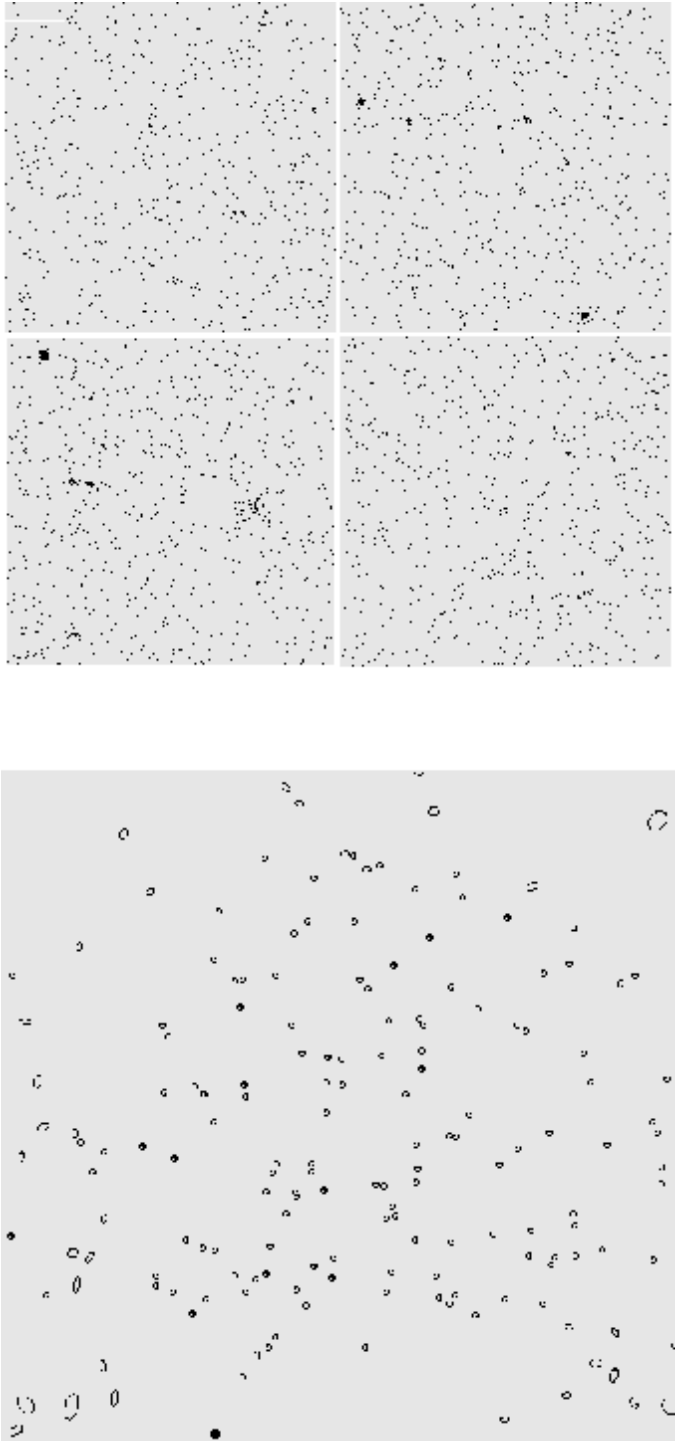}
\end{center}
\caption{
Top: a simulated 30-ksec {\it Chandra} ACIS-I observation of the Lockman Hole.
Data in all four chips are shown, re-binned by a factor of two for 
greater visual clarity.  The gaps between chips are $\approx$ 15 pixels.
Bottom:
ellipses representing the 171 sources detected by our algorithm.  The ellipses,
whose sizes are normally set by deriving the 1$\sigma$ principal axes 
and rotation angle for each source, have minimum axis lengths of 10 
pixels for greater visual clarity.
}
\label{fig:lh_data}
\end{figure}

\clearpage
\begin{figure}
\begin{center}
\includegraphics[angle=-90,scale=0.75]{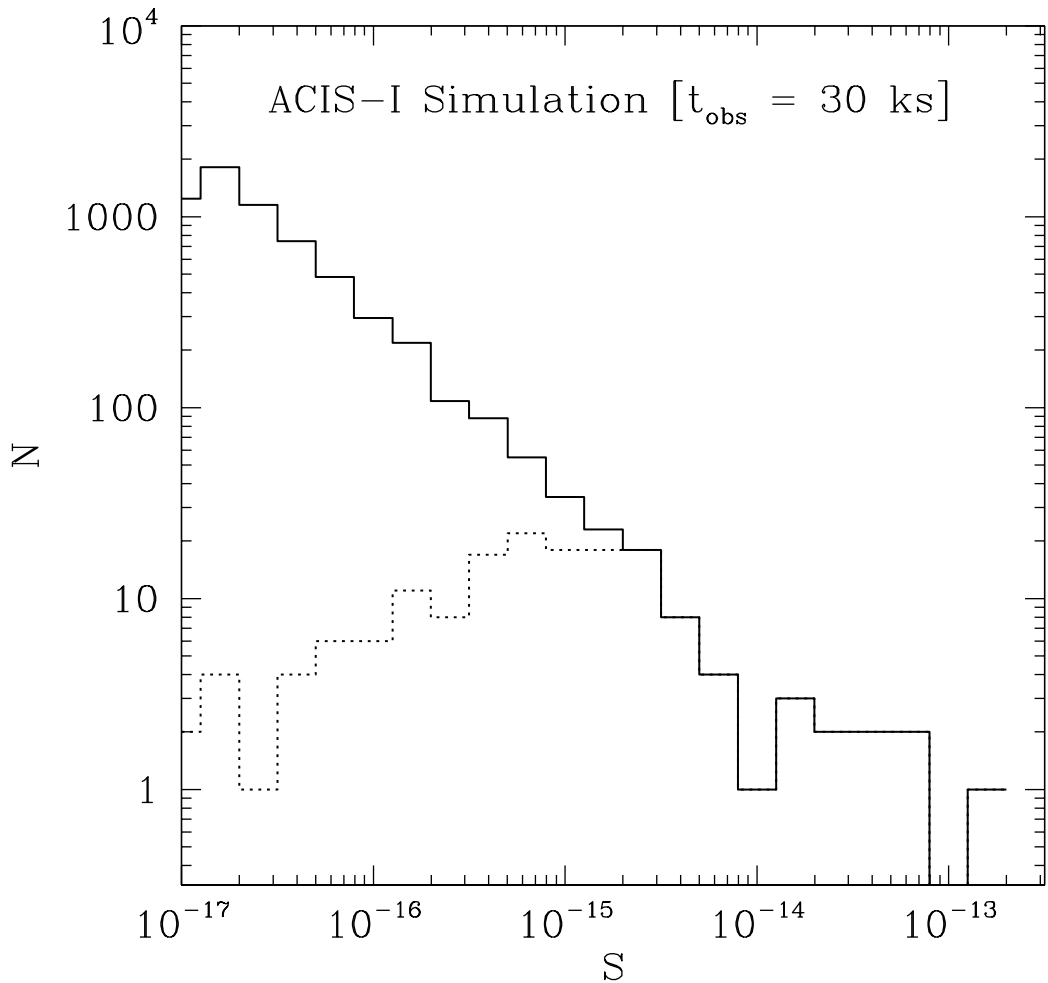}
\end{center}
\caption{
Differential ${\log}N-{\log}S$ distributions
for detected point sources (dotted line) and all point sources (solid
line) in the {\it Chandra} ACIS-I Lockman Hole field.
We conclude that our algorithm will efficiently detect sources
in ACIS-I images with fluxes $\simgreat$ 10$^{-15}$ erg cm$^{-2}$ sec$^{-1}$
(0.5 - 2 keV), and has the ability to detect Poisson fluctuations for
sources with fluxes $\simless$ 10$^{-16}$ erg cm$^{-2}$ sec$^{-1}$.
}
\label{fig:logN}
\end{figure}

\clearpage
\begin{figure}
\begin{center}
\includegraphics[angle=-90,scale=0.75]{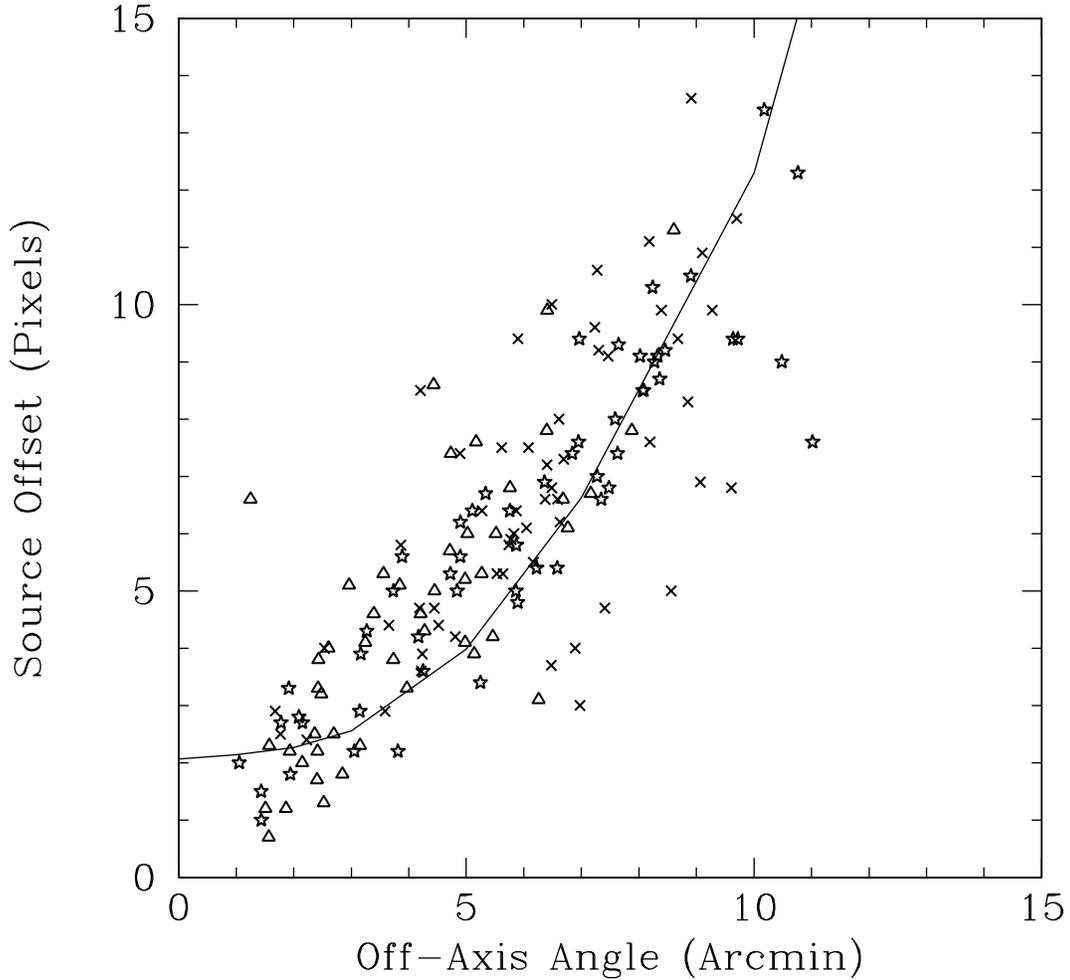}
\end{center}
\caption{
Offsets of the locations of detected sources from their
actual locations within
the {\it Chandra} ACIS-I Lockman Hole field,
as a function of off-axis angle.  These offsets are caused 
by the asymmetry of the {\it Chandra} PSF,
and the solid line represents its 95\% encircled energy radius.
Triangles represent
sources with less than 6 counts; crosses, sources with
between 6 and 15 counts; and stars, sources with more than 15
counts.  
This figure demonstrates that our ability to associate
detected sources with actual sources does not 
vary as a function of source
strength.
}
\label{fig:offax}
\end{figure}

\clearpage
\begin{figure}
\includegraphics[scale=0.75]{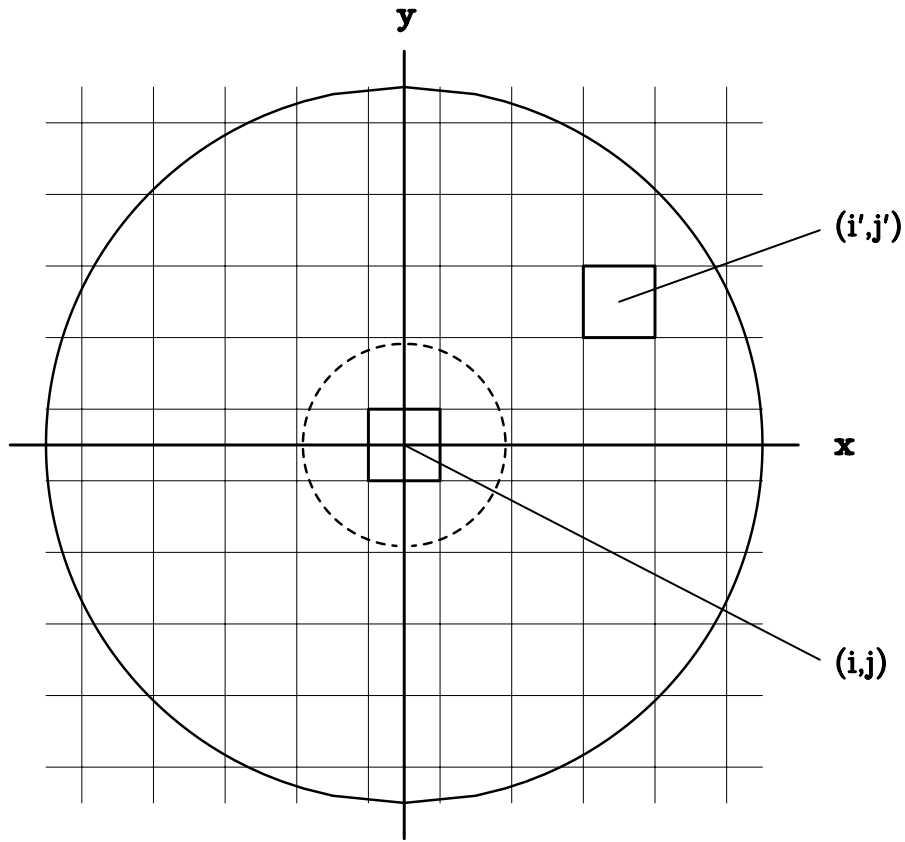}
\caption{
Illustration of variables used when computing
correlation values.  $x$ and $y$ are continuous variables describing the
wavelet function centered in pixel $(i,j)$.  The
correlation value for pixel $(i,j)$ is computed by summing the product of
the data in pixels $(i',j')$ and the integral of the wavelet function 
in those pixels.  The dotted line shows the boundary between the positive
kernel $PW$ and negative annulus $NW$ of a Mexican Hat wavelet with
$\sigma_x = \sigma_y =$ 1 pixel, while the solid line shows the 
extent over which the summation is carried out; beyond this line,
the amplitude of wavelet function is $\approx$ 0.
}
\label{fig:coord}
\end{figure}

\clearpage
\begin{figure}
\begin{center}
\includegraphics[angle=-90,scale=0.60]{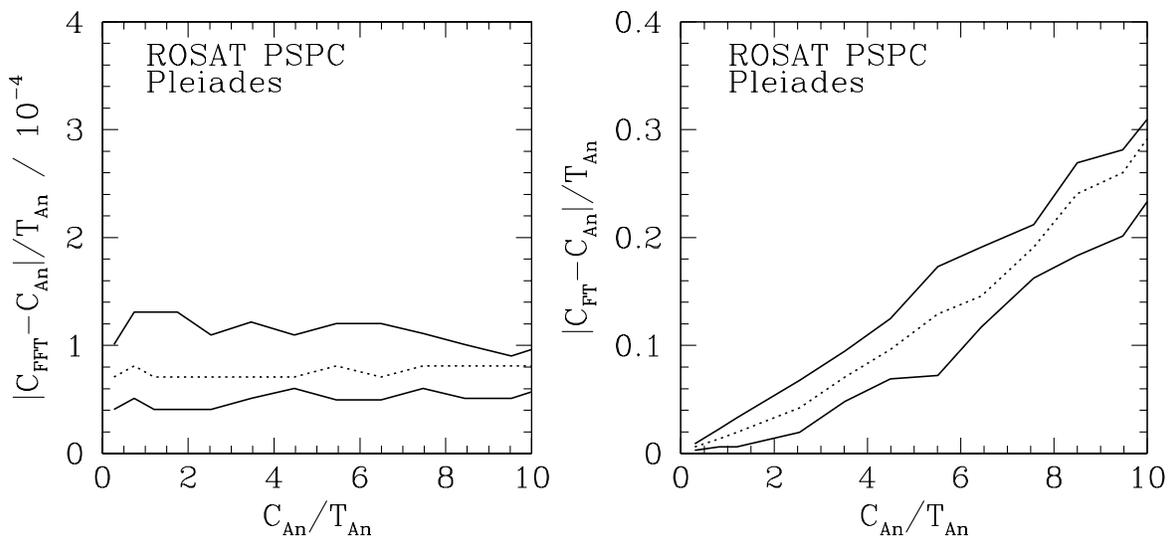}
\end{center}
\caption{
Left: discrepancy between the correlation values determined by
the analytic method and by using the FFT, for the
{\it ROSAT} PSPC data of the Pleiades cluster.  In this
example, $\sigma_x$ = $\sigma_y$ = 2 pixels.
Right: same as left, but comparing the analytic method with the
use of the analytic Fourier Transform for the MH function.
}
\label{fig:ftan}
\end{figure}

\clearpage
\begin{figure}
\begin{center}
\includegraphics[scale=0.60]{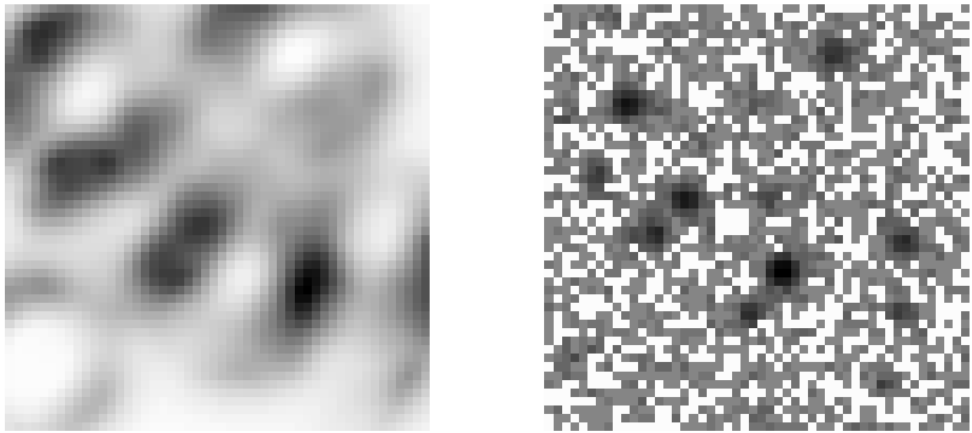}
\end{center}
\caption{
Illustration of the effect of including the computation of covariances
in the calculation of a two-iteration background (see Appendix
\ref{sect:covar} for details).
Left: ratio of variances $V[B_2]_{\rm covar}{\slash}V[B_2]_{\rm no~covar}$,
for a 50$\times$50 subfield of the {\it ROSAT} PSPC Pleiades Cluster image,
analyzed with a wavelet with scale sizes
$\sigma_x$ = $\sigma_y$ = 4 pixels.  Light regions have
values $\approx$ 1, while dark regions have values up to $\approx$ 1.3.
Right: the 50$\times$50 subfield of the {\it ROSAT} PSPC Pleiades image
for which the ratio of variances shown at left was computed.  Sources
in this field correspond with light (low-amplitude) regions in the ratio image.
}
\label{fig:covar}
\end{figure}

\clearpage

\begin{deluxetable}{ll}
\tablenum{1}
\tablecaption{Source Detection Variance Formulae}
\tablewidth{0pt}
\tablehead{
\colhead{Property} & \colhead{Variance}}
\startdata
Correlation ($C_{i,j}$) & $<W^2{\star}D>_{i,j}$ \\
\\
Exposure-Corrected & $V[C_{i,j}]i +$ \\
Correlation ($C_{{\rm cor},i,j}$) & $<W^2{\star}(E^2V[B_{\rm norm}])>_{i,j} + $ \\
 & 2 $E_{i,j} <W^2{\star}(EV[B_{\rm norm}])>_{i,j}$ + \\
 & $E_{i,j}^2 <W^2{\star}V[B_{\rm norm}])>_{i,j}$ \\
\\
Approximate & $V[C_{i,j}] + B_{{\rm norm},i,j} <W^2{\star}E^2>_{i,j}$ \\
Exposure-Corrected & \\
Correlation ($C_{{\rm cor},i,j}^{\rm approx}$) & \\
\\
Normalized Background ($B_{{\rm norm},i,j}$) & $<NW^2{\star}D_N>_{i,j}{\slash}(<NW{\star}E>_{i,j})^2$ \\
\enddata
\tablecomments{$i$ and $j$ are pixel indices, and $N$ is the number of
iterations used to compute the background map, if it is not provided.}
\label{tab:sderr}
\end{deluxetable}

\begin{deluxetable}{lll}
\tablenum{2}
\tablecaption{Source Property Expectation Values}
\tablewidth{0pt}
\tablehead{
\colhead{Property} & \colhead{Units} & \colhead{Expectation Value}}
\startdata
X Location & pix & $x_s~=~(1{\slash}D_s)\sum_{i \in sc}\sum_{j \in sc} D_{i,j} i$ \\
\\
Y Location & pix & $y_s~=~(1{\slash}D_s)\sum_{i \in sc}\sum_{j \in sc} D_{i,j} j$ \\
\\
Bkg Counts & ct & $B_s~=~\sum_{i \in sc}\sum_{j \in sc} E_{i,j}B'_{{\rm norm},i,j}$ \\
\\
Counts & ct & $C_s~=~D_s - B_s$ \\
\\
Exposure & varies & $t_s~=~(t_o{\slash}D_s)\sum_{i \in sc}\sum_{j \in sc} D_{i,j}E_{i,j}$ \\
\\
Count Rate & ct $\slash$ exp unit & $R_{s,C}~=~C_{\rm s}{\slash}t_{\rm s}$ \\
\\
Bkg Count Rate & ct $\slash$ exp unit & $R_{s,B}~=~B_{\rm s}{\slash}t_{\rm s}$ \\
\enddata
\tablecomments{$i$ and $j$ are pixel indices;
$sc$ denotes those image pixels which lie within the source's
cell; $D_s = \sum_{i \in sc}\sum_{j \in sc} D_{i,j}$;
and $t_o$ is the total exposure, in the same units as the exposure map.}
\label{tab:prop}
\end{deluxetable}

\begin{deluxetable}{lll}
\tablenum{3}
\tablecaption{Source Property Variance Formulae}
\tablewidth{0pt}
\tablehead{
\colhead{Property} & \colhead{Units} & \colhead{Variance}}
\startdata
X Location ($x_s$) & pix & $(1{\slash}D_s^2)\sum_{i \in sc}\sum_{j \in sc} D_{i,j} (i-x_s)^2$ \\
\\
Y Location ($y_s$) & pix & $(1{\slash}D_s^2)\sum_{i \in sc}\sum_{j \in sc} D_{i,j} (j-y_s)^2$ \\
\\
Bkg Counts ($B_s$) & ct & $\sum_{i \in sc}\sum_{j \in sc} E_{i,j}^2 V[B'_{{\rm norm},i,j}]$ \\
\\
Counts ($C_s$) & ct & $D_s + V[B_s]$ \\
\\
Exposure ($t_s$) & sec & $(1{\slash}D_s^2)\sum_{i \in sc}\sum_{j \in sc} D_{i,j} ( t_oE_{i,j}-t_s )^2$ \\
\\
Count Rate ($R_{s,C}$) & ct $\slash$ exp unit & $(1{\slash}t_s^2)(V[C_s] + R_{s,C}^2 V[t_s])$ \\
\\
Bkg Count Rate ($R_{s,B}$) & ct $\slash$ exp unit & $(1{\slash}t_s^2)(V[B_s] + R_{s,B}^2 V[t_s])$\\
\enddata
\tablecomments{$i$ and $j$ are pixel indices;
$sc$ denotes those image pixels which lie within the source's
cell; $D_s = \sum_{i \in sc}\sum_{j \in sc} D_{i,j}$;
and $t_o$ is the total exposure, in the same units as the exposure map.
$V[B'_{{\rm norm},i,j}]$ is defined in eq.~(\ref{eqn:corbvar}).}
\label{tab:error}
\end{deluxetable}

\begin{deluxetable}{cccccc}
\tablenum{4}
\tablecaption{Number of Detected Sources: ROSAT PSPC Pleiades Image}
\tablewidth{0pt}
\tablehead{
 & & & \colhead{Scale} & \colhead{Exposure} & \colhead{Number of}\\
\colhead{Test} & \colhead{Iterations} & \colhead{Significance} & \colhead{Separation} & \colhead{Correction} & \colhead{Sources}}
\startdata
1 & 2 & 10$^{-6}$ & 2 & Fast (eq.~\ref{eqn:fastcor}) & 129 \\
2 & 2 & 10$^{-6}$ & $\sqrt{2}$ & Fast & 136 \\
3 & 3 & 10$^{-6}$ & 2 & Fast & 130 \\
4 & 3 & 10$^{-6}$ & $\sqrt{2}$ & Fast & 137 \\
5 & 3 & 10$^{-6}$ & 2 & Full (eq.~\ref{eqn:fullcor}) & 132 \\
6 & 3 & 10$^{-6}$ & $\sqrt{2}$ & Full & 138 \\
7 & 2 & 10$^{-5}$ & 2 & Fast & 144 \\
8 & 2 & 10$^{-5}$ & $\sqrt{2}$ & Fast & 148 \\
\enddata
\tablecomments{Scale sizes range from
$\sigma =$ 1 to $\sigma =$ 16 pixels. 
The threshold significance for source cleansing is
$S_o$ = 10$^{-2}$.}
\label{tab:numdet}
\end{deluxetable}

\begin{deluxetable}{lc}
\tablenum{5}
\tablecaption{Comparison Among Source Detection Algorithms: ROSAT PSPC Pleiades Image}
\tablewidth{0pt}
\tablehead{
\colhead{Method} & \colhead{Number of Detected Sources} }
\startdata
Our Algorithm (Test 8) & 148\\
Damiani et al.~(1997b) & 150\\
Micela et al.~(1996) & 99\\
WGACAT & 127\\
MPE & 100\\
\\
Common To:\\
~~~All Methods& 81\\
~~~Our Algorithm \& WGACAT & 27\\
~~~Our Algorithm \& MPE & 12\\
~~~WGACAT \& MPE & 3\\
~~~Our Algorithm Only & 28 \\
~~~WGACAT Only & 16 \\
~~~MPE Only & 4 \\
\enddata
\label{tab:compare}
\end{deluxetable}


\begin{references}

\reference{} Biviano, A., Durret, F., Gerbal, D., Le Fevre, O., Lobo, C.,
Mazure, A., \& Slezak, E.~1996, \aap, 311, 95

\reference{} Bahcall, J.~N., Flynn, C., Gould, A., \& Kirhakos, S.~1994,
\apj, 435, L51

\reference{} Coupinot, G., Hecquet, J., Auriere, M., \& Futaully, R.~1992,
\aap, 259, 701

\reference{} Damiani, F., Maggio, A., Micela, G., \& Sciortino, S.~1996,
in R\"ontgenstrahlung from the Universe, 
ed. H.~U.~Zimmerman, J.~Tr\"umper, \& H.~Yorke, (MPE Report 263), 679

\reference{} Damiani, F., Maggio, A., Micela, G., \& Sciortino, S.~1997a, 
\apj, 483, 350

\reference{} Damiani, F., Maggio, A., Micela, G., \& Sciortino, S.~1997b, 
\apj, 483, 370

\reference{} Daubechies, I.~1992, Ten Lectures on Wavelets (Philadelphia: SIAM)

\reference{} Eadie, W.~T., Drijard, D., James, F.~E.,
Roos, M., \& Sadoulet, B.~1971, Statistical Methods in Experimental
Physics (Amsterdam: North-Holland)

\reference{} Freeman, P.~E., Kashyap, V., Rosner, R., Nichol, R., Holden, B., 
\& Lamb, D.~Q.~1996, in Astronomical Data Analysis Software and Systems V, 
ed. G.~H.~Jacoby \& J.~Barnes (San Francisco: ASP), 163

\reference{} Freeman, P.~E., Kashyap, V., Rosner, R., \& Lamb, D.~Q.~2001, in
Proceedings of Statistical Challenges in Modern Astronomy III,
ed. G.~J.~Babu \& E.~D.~Feigelson (New York: Springer-Verlag), in press

\reference{} Frigo, M., \& Johnson, S.~G. 1998, Proc. ICASSP 1998, 3, 1381

\reference{} Frisch, P.~C. 1995, SSR, 72, 499

\reference{} Gill, A.~G., \& Henrikson, R.~N.~1990, \apj, 365, L27

\reference{} Gradshteyn, I., \& Ryzhik, I.~1980, Table of Integrals, Series, 
and Products (San Diego: Academic Press)

\reference{} Grebenev, S.~A., Forman, W., Jones, C., \& Murray, S.~1995, 
\apj, 445, 607

\reference{} Harnden, F.~R., Jr., Fabricant, D.~G., Harris, D.~E., \&
Schwarz, J.~1984, SAO Special Report 393

\reference{} Hasinger, G., Johnston, H., \& Verbunt, F.~1994, \aap, 288, 466

\reference{} Hasinger, G., Burg, R., Giacconi, R., Schmidt, M.,
Tr\"umper, J., \& Zamorani, G.~1998, \aap, 329, 482

\reference{} Holschneider, M.~1995, Wavelets: An Analysis Tool (Oxford: Oxford
University Press) 

\reference{} Kashyap, V., Micela, G., Sciortino, S., Harnden, F.~R., Jr., \&
Rosner, R.~1994, in The Soft X-Ray Cosmos, 
ed.~E.~M.~Schlegel, \& R.~Petre (New York: AIP), 239

\reference{} Kashyap, V., Rosner, R., Harnden, F.~R., Jr., Micela, G., \&
Sciortino, S.~1996, in Cool Stars, Stellar Systems, and the Sun,
ed.~R.~Pallavicini, \& A.~K.~Dupree (San Francisco: ASP), 365

\reference{} Langer, W.~D., Wilson, R.~W., \& Anderson, C.~H.~1993,
\apj, 408, L45

\reference{} Lazzati, D., Campana, S., Rosati, P., Chincarini, G., \&
Giacconi, R.~1998, \aap, 331, 41

\reference{} Lazzati, D., \& Chincarini, G.~1998, \aap, 339, 52

\reference{} Mallat, S.~1998, A Wavelet Tour of Signal Processing
(London: Academic Press)

\reference{} Micela, G., Sciortino, S., Kashyap, V., Harnden, F.~R., Jr., \&
Rosner, R.~1996, \apjs, 102, 75

\reference{} Micela, G., et al.~1999, \aap, 341, 751

\reference{} Nichol, R.~C., Holden, B.~P., Romer, A.~K., Ulmer, M.~P.,
Burke, D.~J., \& Collins, C.~A.~1997, \apj, 481, 644

\reference{} Nobile, A., \& Roberto, V.~1986, Comp Phys Comm, 42, 233

\reference{} Nousek, J.~A., \& Lesser, A.~1993, ROSAT Newsletter, 8, 13

\reference{} Pierre, M., \& Starck, J.-L.~1998, \aap, 330, 801

\reference{} Rosati, P., Della Cecai, R., Burg, R., Norman, C., 
\& Giacconi, R.~1995, \apjl, 445, L11

\reference{} Schmidt, M., et al.~1998, \aap, 329, 495

\reference{} Slezak, E., Bijaoui, A., \& Mars, G.~1990, \aap, 227, 301

\reference{} Slezak, E., Durret, F., \& Gerbal, D.~1994, \aj, 108, 1996

\reference{} Snowden, S.~L., \& Kuntz, K.~D.~1998,
Cookbook for Analysis Procedures for ROSAT XRT Observations of Extended
Objects and the Diffuse Background, Part I: Individual Observations
(NASA/GSFC: US ROSAT Science Data Center)

\reference{} Starck, J.-L., Murtagh, F., \& Bijaoui, A.~1995, in
Astronomical Data Analysis Software and Systems IV,
ed.~R.~A.~Shaw, H.~E.~Payne, \& J.~J.~E.~Hayes (San Francisco: ASP), 279

\reference{} Starck, J.-L., \& Murtagh, F.~1998, PASP, 110, 193

\reference{} Starck, J.-L., \& Pierre, M.~1998, A\&AS, 128, 397

\reference{} Stauffer, J.~R., Caillault, J.-P., Gagne, M.,
Prosser, C.~F., \& Hartmann, L.~W.~1994, \apjs, 91, 625

\reference{} Tinney, C.~G., Mould, J.~R., \& Reid, I.~N.~1992, \apj, 396, 173

\reference{} Ulmer, M.~P., Romer, A.~K., Nichol, R.~C., Holden, B.,
Collins, C., \& Burke, D.~1995, BAAS, 187, \#95.03

\reference{} Vikhlinin, A., Forman, W., \& Jones, C.~1994, \apj, 435, 162

\reference{} Vikhlinin, A., McNamara, B.R., Forman, W., Jones, C., Quintana,
H., \& Hornstrup, A.~1998, \apj, 502, 558

\reference{} Voges, W., Gruber, R., Haberl, F., Kuerster, M., Pietsch, W., \&
Zimmermann, U.~1994, ROSAT News 32

\reference{} White, N.~E., Giommi, P., \& Angelini, L.~1994, 
IAU Circ.~6100

\end{references}
\end{document}